\documentclass[12pt,preprint]{aastex}

\newcommand{\be}{\begin{equation}}
\newcommand{\ee}{\end{equation}}
\newcommand{\bex}{\begin{equation}\notag}
\newcommand{\eex}{\end{equation}\notag}
\newcommand{\bea}{\begin{eqnarray}}
\newcommand{\eea}{\end{eqnarray}}
\newcommand{\beax}{\begin{eqnarray*}}
\newcommand{\eeax}{\end{eqnarray*}}
\newcommand{\ba}{\begin{array}}
\newcommand{\ea}{\end{array}}

\newcommand{\vecB}{{\mathbf B}}
\newcommand{\vecE}{{\mathbf E}}
\newcommand{\vecR}{{\mathbf R}}
\newcommand{\vecS}{{\mathbf S}}
\newcommand{\vecA}{{\mathbf A}}

\newcommand{\vecv}{{\mathbf v}}
\newcommand{\vecb}{{\mathbf b}}
\newcommand{\vecr}{{\mathbf r}}

\newcommand{\vecbhat}{{\mathbf {\hat b}}}
\newcommand{\vecxhat}{{\mathbf {\hat x}}}
\newcommand{\vecyhat}{{\mathbf {\hat y}}}
\newcommand{\veczhat}{{\mathbf {\hat z}}}
\newcommand{\scrJ}{{\mathcal{J}}}
\newcommand{\scrB}{{\mathcal{B}}}
\newcommand{\scrF}{{\mathcal{F}}}
\newcommand{\grad}{\mbox{\boldmath$\nabla$}}

\newcommand{\nn}{\nonumber}

\begin{document}

\title{Estimating Electric Fields from Vector Magnetogram Sequences}
\shorttitle{Estimating Electric Fields}


\author{G.~H. Fisher, B.~T. Welsch, W.~P. Abbett, D.~J. Bercik,}
\affil{Space Sciences Laboratory, University of California,
Berkeley, CA 94720-7450}


\begin{abstract}
Determining the electric field distribution on the 
Sun's photosphere is essential for quantitative studies of how energy flows 
from the Sun's photosphere, through the corona, and into the heliosphere. 
This electric field also provides valuable input for data-driven models of the 
solar atmosphere and the Sun-Earth system. We show how
observed vector magnetogram time series can be used to estimate the 
photospheric electric field.  Our method uses a 
``poloidal-toroidal decomposition'' (PTD) of the time derivative of the 
vector magnetic field. These solutions provide
an electric field whose
curl obeys all three components of Faraday's Law.  
The PTD solutions are not unique; the gradient of a 
scalar potential can be added to the PTD electric 
field without affecting consistency 
with Faraday's Law. We then present an iterative technique to determine a 
potential function consistent with ideal MHD evolution; but this
field is 
also not a unique solution to Faraday's Law. Finally, we explore a 
variational approach that minimizes an energy functional to determine a 
unique electric field, a generalization of Longcope's ``Minimum Energy Fit''. 
The PTD technique, the iterative technique, and the variational technique are 
used to estimate electric fields from a pair of synthetic vector 
magnetograms taken 
from an MHD simulation; and these fields are compared with the 
simulation's known electric fields. 
The PTD and iteration techniques compare favorably to results from existing
velocity inversion techniques.
These three techniques are then applied 
to a pair of vector magnetograms of solar active region NOAA AR8210, 
to demonstrate the methods with real data. 

Careful examination of the results from all three methods indicates that
evolution of the magnetic vector by itself does not provide enough
information to determine the true electric field in the photosphere.
Either more information from other measurements, or physical constraints
other than those considered here are necessary to find the true electric
field.
However, we show it is possible to construct physically reasonable
electric field distributions whose curl matches the evolution of 
all three components
of $\vecB$. We also show that the horizontal and vertical Poynting flux 
patterns derived
from the three techniques are similar to one another for 
the cases investigated.
\end{abstract}


\section{Introduction}
\label{section:intro}
\noindent
(A version of this manuscript with higher quality images is at
\url{http://tinyurl.com/yzxn922}).

The availability of frequent, high quality photospheric vector
magnetogram observations from ground-based instruments such as SOLIS
\citep{Henney2009}, space-based instruments such as the 
Hinode/SOT SP \citep{Tsuneta2008} and the planned HMI
instrument on SDO \citep{Scherrer2005} motivates a fresh
look at how these data can be used for quantitative studies of the
dynamic solar magnetic field.  

While the reduction of the observed 
polarimetry data into maps of the three magnetic
field components is itself a challenging problem, we assume for simplicity
in this paper that that problem has been solved, and that time 
sequences of error-free vector magnetic field maps have been
obtained over some arbitrary field of view at the
solar photosphere.  We will not address the implications of errors
in the measurements, problems with data reduction, 
or uncertainties such as the resolution of the 180 degree ambiguity.

There are many possible uses of vector magnetic field maps of the photosphere.  
We will focus on just one: the use of time
sequences of vector magnetograms to determine the surface distribution
of the electric field vector on the Sun.

Attempts to measure electric fields on the Sun have been
made using spectropolarimetric techniques designed to measure the linear
Stark effect in \ion{H}{1} spectral lines \citep{MoranFoukal1991a}.  Attempts to
measure the electric field in prominences showed no signal
above the measurement threshold, but these techniques applied to
a small flare surge did result in a measurement above the
threshold \citep{FoukalBehr1995}.  In this paper, we attempt
to determine the electric field in the solar atmosphere
indirectly from vector magnetic field measurements, using Faraday's Law, rather
than appealing to the Stark effect.

There are several reasons why determining the electric field on the
solar surface is useful.  The simultaneous determination
of three-component magnetic and electric field vectors will allow us
to estimate the Poynting flux of electromagnetic energy entering
the corona, as well as the flux of relative magnetic helicity.  In
addition, under the assumptions of ideal MHD, where $\vecE = -\vecv
/c \times \vecB$, it allows us to estimate the three-component
flow field in the photosphere.  Knowing either the flows or an
electric field consistent with Faraday's Law will enable
the driving of MHD models of the solar atmosphere that are consistent
with observed data.  This is a key requirement for
predictive, physics-based models of the solar atmosphere that might be
used in forecasting applications.

\begin{figure}
\includegraphics[width=6.5in]{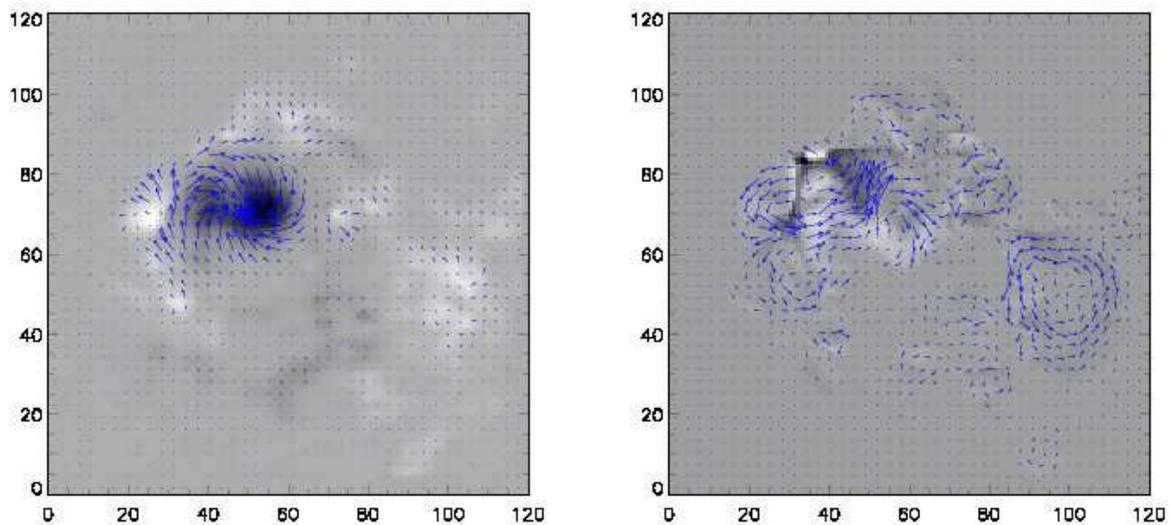}
\caption{
The left panel shows a vector magnetogram of Active Region 8210 taken with
the University of Hawaii's IVM instrument.  This dataset is
described in \cite{Welsch2004}.  Arrows show the directions
and amplitudes of $B_x$ and $B_y$, and the background image shows the
amplitude of $B_z$.  The right panel shows a ``vector electrogram''
(a three-dimensional vector electric field map) of the active region,
using the time evolution of $\vecB$ to estimate $\vecE$.  
Arrows show
estimated directions and amplitudes of $E_x$ and $E_y$, while the background
image shows the estimated amplitude of $E_z$.  
The example shown
here displays $\vecE$ computed using the ``variational'' technique
(\S \ref{subsection:variational}).  A detailed discussion of the calculation 
is in \S \ref{section:8210}.
}
\label{figure:intro}
\end{figure}

Most recent research on deriving electric fields in the solar
atmosphere has been done by either explicitly or implicitly invoking
the ideal MHD assumption, $\vecE = -\vecv /c \times \vecB$, and has
focused on deriving two- or three-component flow fields from time
sequences of magnetograms.  Such techniques can be divided into two
classes, which we call ``tracking methods'', and ``inductive
methods''.
%
%

Tracking methods, such as the local correlation tracking (LCT)
approach, first developed by \citet{November1988}, find a velocity
vector by computing a cross-correlation function that depends on the
shift between sub-images or tiles when comparing two images.
The shift that maximizes the cross-correlation function
(or alternatively, minimizes an error function) is then
taken to be the displacement between the two sub-images; this
displacement divided by the time between the images is then
defined as the average local velocity.  The velocity field over the
entire image is built up by repeating this process for all image
locations.
%
%
The LCT technique suffers from two shortcomings: (1) the
technique does not assume any physical conservation laws, meaning that
the derived velocity field may not have a physical connection to the
real flow field in the solar atmosphere; and (2) the technique is
intrinsically two-dimensional, and does not account for vertical flows
or evolving three-dimensional structures in the solar atmosphere.  On the
other hand, LCT techniques offer the advantage of being simple and
robust, and are able to use non-magnetic data, such as white-light
or G-band images for estimating flow fields, though the results must
then be interpreted carefully.  Examples of LCT techniques in current use
in the solar research community include the implementation by
\citet{November1988}, FLCT \citep{Fisher2008}, and Lockheed-Martin's
LCT code \citep{Title1995,Hurlburt1995}.

Inductive methods of flow inversion from magnetograms were pioneered
by \citet{Kusano2002}, who used a combination of horizontal flow
velocities derived from LCT techniques applied to the normal component
of vector magnetograms, along with a solution to the vertical
component of the magnetic induction equation, to derive a
three-component flow field from a sequence of vector magnetograms.  An
alternate approach, using the same idea, but adding the
interpretation of \citet{Demoulin2003} plus a Helmholtz decomposition
of the ``flux transport velocity'' was proposed by
\citet{Welsch2004}.  \citet{Longcope2004} combined the vertical
component of the induction equation with a variational
constraint that minimizes the total kinetic energy of the photosphere 
while still obeying the normal component of the
induction equation.  Additional
techniques minimize a localized error functional
while ensuring that the vertical component of the
induction equation is satisfied \citep{Schuck2006, Schuck2008}.

\citet{Kusano2002} observed that the $\partial B_x /\partial t$ and
$\partial B_y / \partial t$ components of the induction equation
involve vertical derivatives of horizontal electric field terms
($\partial E_y/ \partial z$ and $\partial E_x/ \partial z$,
respectively).  They noted that these electric field components are
unconstrained by single-height vector magnetogram sequences.
Consequently, while the two horizontal components of the induction
equation provide more information about the evolution of the vector
magnetic field, they introduce two more unknowns to the system.

The primary goal of this paper is to explore the extent to which
an electric field consistent with the evolution of {\em all three
components} of $\vecB$ can be derived from a sequence of
single-height vector magnetograms, despite incomplete information
about $\partial E_x/ \partial z$ and $\partial E_y/ \partial z$.
We first show that it is possible to derive an electric field whose
curl is equal to the time derivatives of all three components
of $\vecB$.
However, this electric field is not unique, and knowledge of
additional physics of the electric field formation is necessary to
further constrain the solution.
We then show that a solution for the electric field, under the
assumptions of ideal MHD, can be derived through the
use of an iterative procedure, or alternatively, as the
solution of a variational problem.
We will compare and contrast these solutions with results from an MHD
simulation, where the electric field
is known.  To illustrate this methodology with real data,
we then apply these techniques to a set of vector
magnetograms of NOAA AR8210 from 1 May 1998.  Figure \ref{figure:intro} shows
a vector electric field map (a ``vector electrogram'') of NOAA AR8210,
along with a vector magnetogram of the same active region, derived
using one of the techniques discussed in this paper.
%


The remainder of this paper is structured as follows.  In \S
\ref{section:ptd}, we derive solutions to the electric field given
the vector magnetic field evolution in a single horizontal
plane.  These solutions use the poloidal-toroidal
\citep{Chandrasekhar1961, Moffatt1978} decomposition of the
electric field.  

In \S \ref{section:psi} we show how additional physics
describing the electric field can be included 
by exploring two approaches, one iterative and the other
variational, both aiming for the construction of an ideal MHD electric
field consistent with $\vecE = - \vecv /c \times \vecB$.  While the
ideal MHD assumption is believed to be a good approximation in
the photosphere, the variational formalism has been generalized
to include non-ideal effects.  In \S \ref{section:psi} we also compare
and constrast electric fields derived with our techniques to a
test case where the true solutions are known, and in
\S \ref{section:8210} with an example using
real vector magnetic field data.
We discuss and summarize our results in \S \ref{section:conclusions}.

\section{Poloidal-Toroidal Decomposition}
\label{section:ptd}

\subsection{Decomposing the Magnetic Field and its Time Derivative}
\label{subsection:bdecomp}
The poloidal-toroidal decomposition (henceforth PTD) of the magnetic
field into two scalar potentials is well-known among dynamo theorists
\citep{Moffatt1978} and has been used extensively in MHD models of the
solar interior that employ the anelastic approximation
\citep{Glatzmaier1984,
%
%
Lantz1999,Fan1999,Brun2004}.  The formalism appears to have been introduced by
\cite{Chandrasekhar1961}.  
%
%
%
Here, we will briefly discuss the PTD of the magnetic field, but will
focus most of our effort on the PTD of the partial
time derivative of the magnetic field, because of its connection to 
the electric field.  Here we use $\vecB$ to refer to a snapshot
of the magnetic field within the vector magnetogram field of view, and
$\dot \vecB$ to refer to its partial time derivative.

Given a snapshot of the three-component magnetic field distribution in Cartesian
coordinates, one can write $\vecB$ as follows:
\be
\vecB = \grad \times \grad \times \scrB \veczhat + \grad \times \scrJ \veczhat .
\label{equation:ptdb}
\ee
Here, $\scrJ$ is referred to as the ``toroidal'' potential, and $\scrB$ as
the ``poloidal'' potential. 
The vector
potential $\vecA$ is then given by
\be
\vecA = \grad \times \scrB \veczhat + \scrJ \veczhat + \grad \xi ,
\label{equation:vecp}
\ee
where $\xi$ is a gauge potential, left unspecified.
%

Taking the partial time derivative of equation (\ref{equation:ptdb})
leads to an equation of exactly the
same form 
for $\dot \vecB$ in terms of the partial time derivatives $\dot \scrB$ and
$\dot \scrJ$:
\be
\dot \vecB = \grad \times \grad \times \dot \scrB \veczhat + \grad \times \dot 
\scrJ \veczhat\ .
\label{equation:ptdbdot}
\ee

The vector $\veczhat$ is assumed to
point in the vertical direction, $i.e.$ normal to the photosphere.  
A subscript $z$ will denote vector components in the $\veczhat$ direction, and
a subscript $h$ will denote vector components or derivatives in the locally
horizontal directions, parallel to the tangent plane of the photosphere.

It is useful to rewrite equations (\ref{equation:ptdb}) 
and (\ref{equation:ptdbdot}) in terms of 
horizontal and vertical derivatives as
\be
\vecB = \grad_h \left( \partial \scrB \over \partial z \right) 
+ \grad_h \times \scrJ \veczhat
- \nabla_h^2 \scrB \veczhat ,
\label{equation:ptdbexpand}
\ee
and
\be
\dot \vecB = \grad_h \left( \partial \dot \scrB \over \partial z \right) 
+ \grad_h \times \dot \scrJ \veczhat
- \nabla_h^2 \dot \scrB \veczhat . 
\label{equation:ptdbdotexpand}
\ee

The PTD of $\dot \vecB$ has a useful connection to
observation.  Examining only the z-component of
equation (\ref{equation:ptdbdotexpand}), one finds
\be
\nabla_h^2 \dot \scrB = -\dot B_z\ ,
\label{equation:poissonbz}
\ee
where $\dot B_z$ is the partial time derivative of the vertical component
of the magnetic field,
and $\nabla_h^2$ is the horizontal contribution to the Laplacian.
Thus knowledge of $\dot B_z$ in a layer yields a solution for 
$\dot \scrB$
by solving a horizontal, two-dimensional Poisson equation.

Taking the curl of equation (\ref{equation:ptdbdotexpand}) 
and examining only the z-component of the result, one finds
\be
\nabla_h^2 \dot \scrJ = -( 4 \pi / c ) \dot J_z = - \veczhat 
\cdot ( \grad \times \dot \vecB_h ) .
\label{equation:poissonjz}
\ee
Knowing the time derivative of the horizontal field
$\dot \vecB_h$ in a layer, and
hence the vertical component of its curl, determines $\dot \scrJ$
in that layer, from solutions to another Poisson equation.

Finally, taking the horizontal divergence of equation
(\ref{equation:ptdbdotexpand}) results in
\be
\nabla_h^2 ( {\partial \dot \scrB / \partial z} ) = \grad_h \cdot \dot \vecB_h .
\label{equation:poissondivbh}
\ee
Here, knowing the horizontal divergence of 
$\dot \vecB_h$ allows one to determine 
$\partial \dot \scrB / \partial z$, 
once again by solving a two-dimensional Poisson equation.

It is worth noting an additional implication of equation
(\ref{equation:poissondivbh}).  From the solenoidal nature of $\vecB$
it follows that
\be
\grad_h \cdot \dot \vecB_h = - \partial \dot B_z / \partial z\ .
\label{solenoidal}
\ee
Thus equation (\ref{equation:poissondivbh}) can be regarded as the
partial z-derivative of equation (\ref{equation:poissonbz}), yet no
depth derivatives of the data were needed to evaluate it.  We return
to this point later.

To find the PTD of the magnetic field itself rather than its time derivative, 
the quantities $\scrB$, $\scrJ$,
and $\partial \scrB / \partial z$ obey exactly the same Poisson equations
(\ref{equation:poissonbz}-\ref{equation:poissondivbh}) above, 
but with all of the overdots in the equations removed.
This description of $\vecB$ allows
for an alternate computation of
potential magnetic fields valid at
the vector magnetogram surface 
(Appendix \ref{app:a}), in which the 
divergence of $\vecB_h$ taken from the vector magnetogram
can be incorporated into the solution.  
This formulation for the potential
magnetic field has applications for: 
({\it i}) estimating the flux of free magnetic
energy, $e.g.$, equation (23) of \cite{Welsch2006}, where one needs to
subtract the measured and the potential-field values of the horizontal magnetic
field, and ({\it ii}) computing the vector-potential $\vecA_P$
of the potential field, useful in estimating the magnetic helicity flux.
The horizontal magnetic field can be decomposed into
the potential-field contribution $\grad_h ( \partial \scrB / \partial z )$
and the non-potential contribution 
$\vecB ^{\scrJ}_h = \grad_h \times \scrJ \veczhat$ (see Appendix \ref{app:a}).
Welsch's flux of free
energy then becomes 
\be
S^{\rm free}_z = {1 \over 4 \pi} c \vecE_h \times \vecB^{\scrJ}_h\ ,
\label{equation:freeenergy}
\ee
where methods of estimating $\vecE$ will be discussed below.

\subsection{Finding an Electric Field from Faraday's Law}
\label{subsection:ptde}

Now, compare
equations (\ref{equation:ptdbdot}) 
and (\ref{equation:ptdbdotexpand}) with 
Faraday's law relating the time derivative of $\vecB$ to the curl of the
electric field:
\be
\dot { \vecB }  = -c \grad \times \vecE
\label{equation:faraday}
\ee
Equating the expressions for $\dot { \vecB}$, 
%
%
one finds this expression for $c \grad \times \vecE$:
%
%
\bea
c \grad \times \vecE 
&=& - \grad \times \grad \times \dot \scrB \veczhat - \grad \times \dot \scrJ 
\veczhat
\label{equation:curleptd} \\
&=& - \grad_h (\partial \dot \scrB / \partial z ) - 
\grad_h \times \dot \scrJ \veczhat + \nabla_h^2 \dot \scrB \veczhat.
\label{equation:curleptdexpand}
\eea
Uncurling equation (\ref{equation:curleptd}) yields this expression for the
electric field $\vecE$:
\be
c \vecE = - \grad \times \dot \scrB \veczhat - \dot \scrJ \veczhat - 
c \grad \psi \equiv c \vecE^I - c \grad \psi .
\label{equation:eptd}
\ee
In the process of uncurling equation (\ref{equation:curleptd}), it is
necessary to add the (three-dimensional) gradient of an unspecified
scalar potential $\psi$ to the expression for the total electric
field.  The potential $\psi$ can be equated to $- \dot \xi / c + \lambda$, 
where $\xi$
is the gauge potential from equation (\ref{equation:vecp}), and $\lambda$ is
some other potential function.  However, we simply let
the electric field 
potential $\psi$ absorb both contributions, and will not use
the gauge $\xi$.
The part of $\vecE$ without the contribution
from $-\grad \psi$ will be henceforth denoted $\vecE^I$, the purely inductive
contribution to the electric field.

We have derived an expression for $\grad \times
\vecE$, including the two horizontal components of the
induction equation, simply by using time derivative information
contained within a single layer, even though those two components of
the induction equation include vertical derivatives of $E_x$ and $E_y$.
This was made possible by equation 
(\ref{equation:poissondivbh}),
which enabled the evaluation of the needed depth derivatives through
the relation $\grad \cdot \dot { \vecB } = 0$.


\subsection{Boundary Conditions}
\label{subsection:boundaries}

To solve the three Poisson equations (\ref{equation:poissonbz}), 
(\ref{equation:poissonjz}), and (\ref{equation:poissondivbh}), 
one must consider their boundary conditions.
For many of the MHD simulation test cases we have used,
the simulation vector fields
obey periodic boundary conditions, which makes the problem straightforward:
one can use Fast Fourier Transform (FFT) 
techniques to solve the Poisson equations without special
consideration for boundary conditions
(see Appendix \ref{app:b}).
For an arbitrary vector magnetogram taken over a finite area, however,
the magnetic field will generally not be periodic, but will be
determined by the measured fields on the boundary.
For equations (\ref{equation:poissonjz}) and (\ref{equation:poissondivbh}), the
horizontal components of $\dot \vecB$ determine the 
boundary conditions
for $\dot \scrJ$ and 
$\partial \dot \scrB / \partial z$
from the x and y components of the
primitive equation (\ref{equation:ptdbdotexpand}):
\be
\dot B_x = {\partial \over \partial x} {\partial \dot \scrB \over \partial z} +
{\partial \dot \scrJ \over \partial y}, 
\label{equation:bxeqn}
\ee
and
\be
\dot B_y = {\partial \over \partial y} {\partial \dot \scrB \over \partial z} -
{\partial \dot \scrJ \over \partial x} \ .
\label{equation:byeqn}
\ee
From these equations, coupled Neumann boundary conditions can be derived:
\be
{\partial \over \partial n}{\partial \dot \scrB \over \partial z} = \dot B_n
-{\partial \dot \scrJ \over \partial s} ,
\label{equation:bdrydbdz}
\ee
and
\be
{\partial \dot \scrJ \over \partial n} = - \dot B_s + 
{\partial \over \partial s}
{\partial \dot \scrB \over \partial z}\  ,
\label{equation:bdryj}
\ee
where subscript $n$ denotes components or derivatives in the direction
of the outward normal from the boundary of the magnetogram, and subscript
$s$ denotes components or derivatives in the counter-clockwise
direction along the magnetogram boundary.

The choice of boundary conditions for $\dot \scrB$ in equation
(\ref{equation:poissonbz}) can have subtle effects on the solution
$\dot B_z$.  
If homogeneous Neumann boundary conditions (derivative normal to the
magnetogram boundary specified with zero slope)
are applied to $\dot \scrB$, then the horizontal curl of $\vecE_h^I$ around
the magnetogram boundary necessarily vanishes, meaning
the average value of 
$\dot B_z$ within the magnetogram
is forced to zero.
If the average value
of $\dot B_z$ is nonzero, the solutions will not reflect this.
This problem can be corrected {\it post facto}, 
however, by adding a correction term to the electric
field -- see Appendix \ref{app:c}.

Applying homogenous Dirichlet boundary conditions (setting 
$\dot \scrB$ to zero at the boundaries) for the solution of
equation (\ref{equation:poissonbz}) for $\dot \scrB$ 
can also result in artifacts if care
is not taken.  Homogenous Dirichlet boundary conditions on $\dot \scrB$ 
imply no net change in $\dot \scrB$ across the magnetogram, resulting
in average $x-$ and $y-$
components of the electric field $\vecE_h^I$ that 
are zero.  Some evolutionary patterns,
such as the emergence and separation of a simple magnetic bipole
oriented in the $x-$ direction should result in a non-zero average value of
$E_y^I$, because the opposite polarities have opposite velocities.  
Thus, if Dirichlet boundary conditions are used in the solution of
equation (\ref{equation:poissonbz}) for $\dot \scrB$, 
some other technique must be used to
find the magnetogram-averaged values of $E_x^I$ and $E_y^I$.


The coupled boundary conditions
(\ref{equation:bdrydbdz}-\ref{equation:bdryj}) for the two Poisson
equations (\ref{equation:poissonjz}-\ref{equation:poissondivbh}) are
degenerate, in that there is a family of coupled non-zero solutions to
the homogeneous Cauchy-Riemann equations for $\dot \scrJ$ and 
$\partial \dot \scrB / \partial z$: 
\be
{\partial \over \partial x} \left( {\partial \dot \scrB \over \partial z} 
\right) 
= - {\partial \dot \scrJ \over \partial y} ,
\label{equation:cauchybeta}
\ee
and
\be
{\partial \over \partial y} \left( {\partial \dot \scrB \over \partial z} 
\right) = 
{\partial \dot \scrJ \over \partial x}
\label{equation:cauchyj}
\ee
which satisfy the boundary conditions for zero time derivative
of the horizontal magnetic field on the boundary. 
These solutions can be added to solutions for $\dot \scrJ$ and 
$\partial \dot \scrB / \partial z$
without changing the 
time derivative of the horizontal
field on the
boundary.  Solutions to equations 
(\ref{equation:cauchybeta}-\ref{equation:cauchyj}) 
also are harmonic, $i.e.$ each of the two solutions also 
obey the two-dimensional (horizontal) Laplace's equation.
%

The boundary conditions needed to solve the Poisson equations for
$\scrB$, $\scrJ$, and $\partial \scrB / \partial z$ (if one needs to
perform the PTD on $\vecB$, rather than $\dot \vecB$) are identical
to the boundary conditions described above, but with all of the overdots
removed.  The Cauchy Riemann degeneracy described above also applies to
the solutions of $\scrJ$ and $\partial \scrB / \partial z$.  

%
%

\begin{figure}
\includegraphics[width=6.5in]{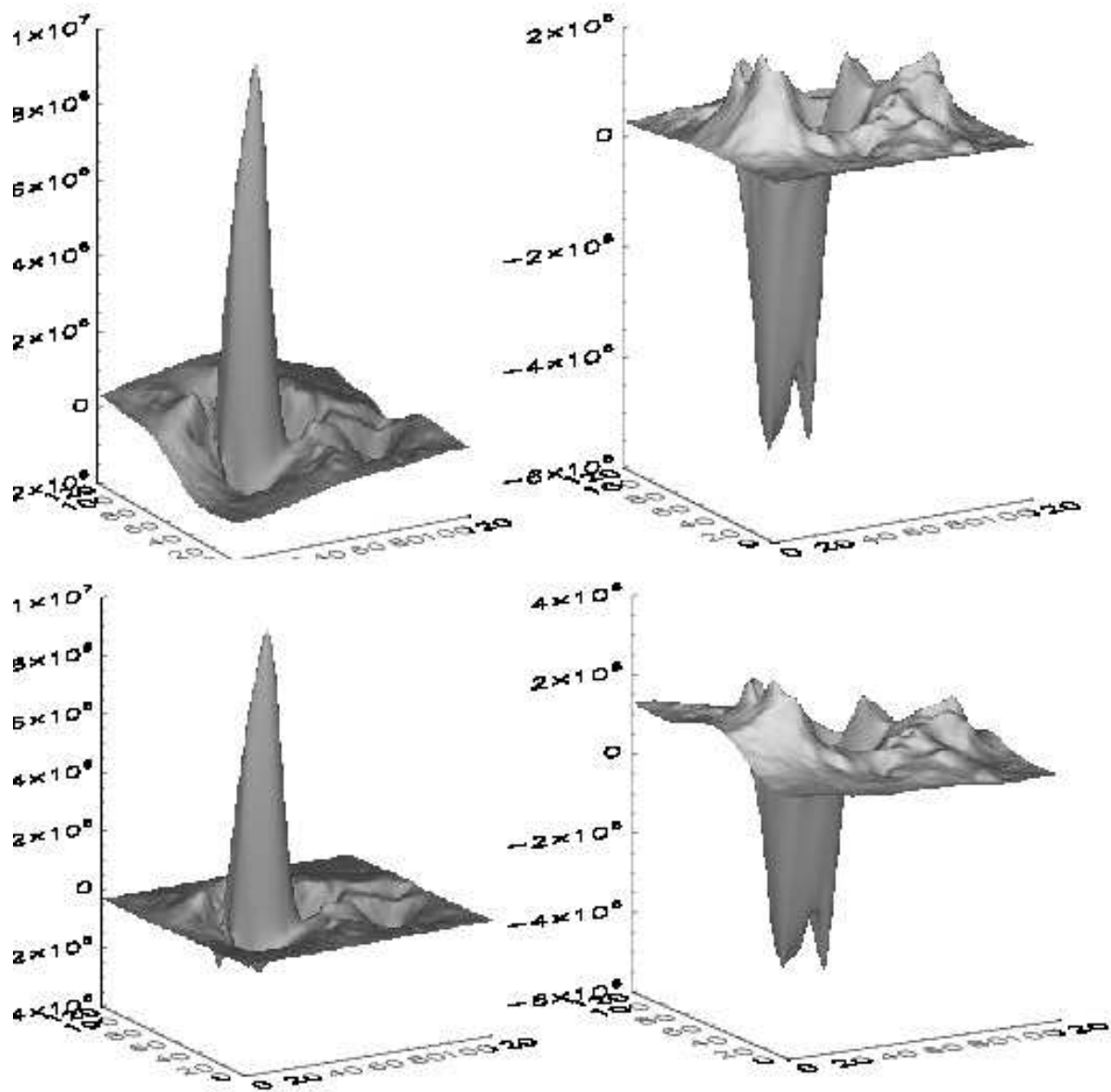}
\caption{
The top two panels show Poisson equation solutions for 
$\partial \scrB / \partial z$ (left) and $\scrJ$ (right) 
within the AR 8210 magnetogram field-of-view, under the assumption that 
$\partial \scrJ / \partial s = 0$ along the magnetogram
boundary.  The AR8210 vector magnetogram data is discussed further in
\S \ref{section:8210}.
The bottom two panels show Poisson equation solutions for
the same two functions under the assumption that 
$\partial / \partial s ( \partial \scrB / \partial z ) = 0$.  The difference
between the upper and lower sets of solutions obey the homogeneous 
Cauchy-Riemann equations.  The two different sets of functions yield identical
values for the horizontal components of the magnetic field.
}
\label{figure:cauchy}
\end{figure}

Even without considering $\grad \psi$, the degeneracy
of the solutions for $\dot \scrJ$ and $\partial \dot \scrB / \partial z$ for
non-periodic boundary conditions means
that particular solutions of the homogeneous
Cauchy-Riemann equations can be added to the solutions for
the inductive electric field $c \vecE^I$ without affecting $\grad \times \vecE$.
This means there is some freedom to specify the solution of
$\dot \scrJ$ or $\partial \dot \scrB / \partial z$ at the boundary.
This freedom cannot be applied to both functions simultaneously, however,
since the two solutions are coupled together.  In practice, this
means that one could choose to set the derivatives parallel to the boundary
of {\it one} of these two functions to zero, 
for example, while still obeying the coupled boundary conditions 
(\ref{equation:bdrydbdz}) and (\ref{equation:bdryj}).  For
an illustration of this degeneracy applied to the PTD solutions
for $\vecB$, 
see Figure \ref{figure:cauchy}.

When periodic boundary conditions cannot be assumed,
equation (\ref{equation:eptd}) shows that the electric field
on the boundary of the magnetogram determines the boundary conditions for
$\dot \scrB$ and $\psi$: 
\be
{\partial \dot \scrB \over \partial n} = c E_s + c{\partial \psi \over \partial s},
\label{equation:bdrybetadot}
\ee
and
\be
c {\partial \psi \over \partial n} = - c E_n - 
{\partial \dot \scrB \over \partial s} .
\label{equation:bdrypsi}
\ee
These coupled Neumann boundary conditions have a similar form
as those for $\dot \scrJ$ and 
$\partial \dot \scrB / \partial z$, 
(equations [\ref{equation:bdrydbdz}-\ref{equation:bdryj}])
except that they depend on horizontal electric fields
rather than on time derivatives of the horizontal magnetic fields.
The practical difficulty with using equations 
(\ref{equation:bdrybetadot}-\ref{equation:bdrypsi}) is that generally one
does not know the behavior of the horizontal electric field vector on the
boundaries if the boundaries are in regions of strong, evolving
magnetic field .  On the other hand,
if one can take the boundaries along regions with weak
average magnetic field strength, one could probably set $E_n$ and $E_s$
to zero along the boundaries.

%

\section{Determining the Scalar Potential}
\label{section:psi}

\subsection{The Implications of Not Specifying $\psi$}
\label{subsection:nopsi}

We can test our approach by applying equations
(\ref{equation:curleptd}) and (\ref{equation:eptd}) to the magnetic
evolution sampled from a thin slab of an MHD simulation, in which
$\vecE$ and $\grad \times \vecE$ are both known.  How well do the
derived results compare with the known electric field, and with the
curl of that electric field?

To address this question, we use the results from the ANMHD
simulations described in \cite{Welsch2007}, which have been used for several
studies of velocity field inversions
\citep{Welsch2008a,Schuck2008}.  This simulation models a magnetic
bipole emerging through a strongly convecting layer.
To compute the reconstructed distribution of $\grad \times \vecE$
and $\vecE$ using PTD, we use time differences from two consecutive
output steps from the MHD simulation to find estimates of the time
derivative of the three components of $\vecB$.  From these three
components, the three Poisson equations
(\ref{equation:poissonbz}-\ref{equation:poissondivbh}) can be
solved, subject to boundary conditions
(\ref{equation:bdrydbdz}-\ref{equation:bdryj}) for $\partial
\dot \scrB / \partial z$ and $\dot \scrJ$.  To solve the three Poisson equations
along with these coupled boundary conditions, we first use a simple
successive over-relaxation method to solve the equation for $\dot
\scrB$, assuming homogenous Neumann (zero gradient) boundary conditions.
We choose Neumann over Dirichlet boundary conditions because Dirichlet
boundary conditions on $\dot \scrB$ lead to zero average horizontal
electric field within the magnetogram 
(see earlier discussion in \S \ref{subsection:boundaries}).
To solve the two Poisson equations
(\ref{equation:poissonjz}-\ref{equation:poissondivbh}) along with
their coupled Neumann boundary conditions
(equations [\ref{equation:bdrydbdz}] and [\ref{equation:bdryj}]), 
we have
adapted the Newton-Krylov solver from RADMHD \citep{Abbett2007} to solve this
elliptic system simultaneously.
%
%
%

\begin{figure}
\includegraphics[width=5.5in]{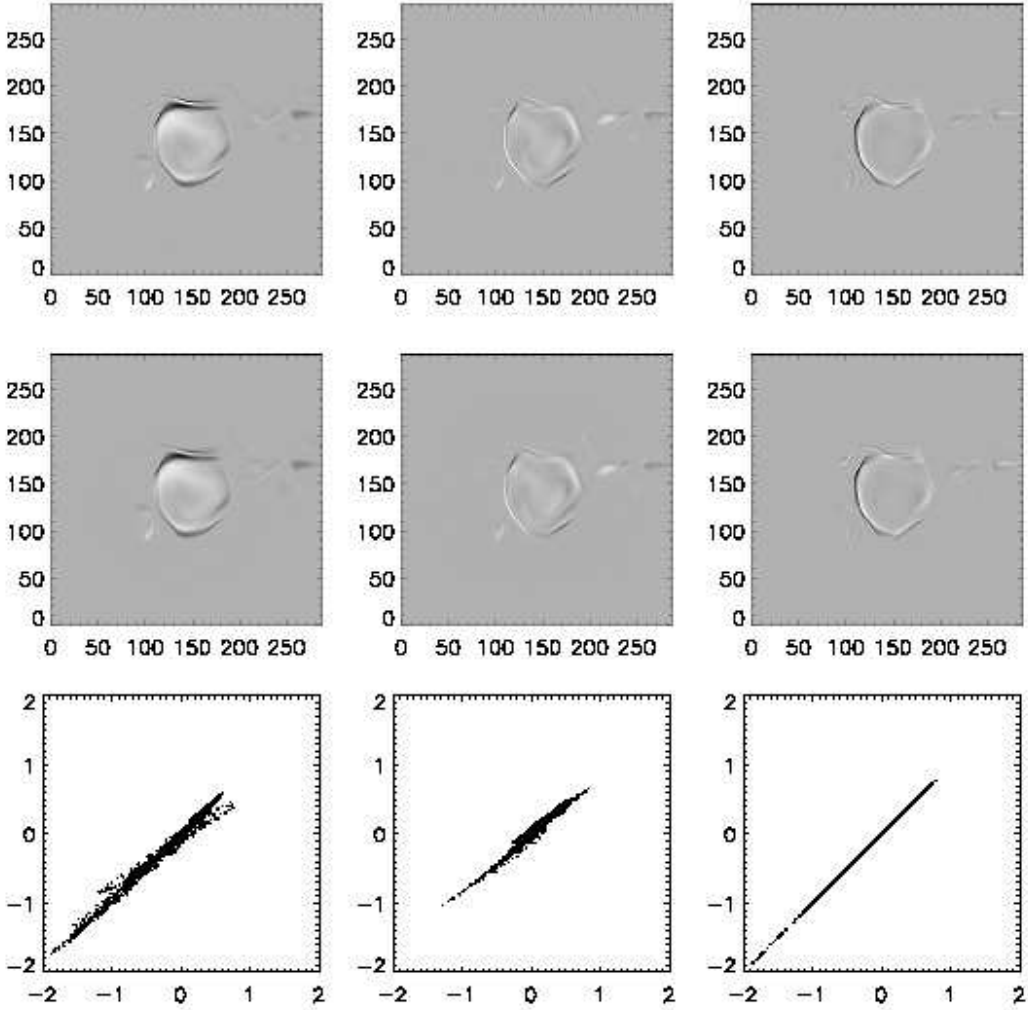}
\caption{
Top panels show the values of the $x-$, $y-$, and $z-$ components of
$c \grad \times \vecE = - \partial \vecB / \partial t$ taken from the ANMHD
simulations described in the text.  The middle panels show the same components
of $c \grad \times \vecE$ determined with the PTD formalism 
(equations [\ref{equation:curleptd}-\ref{equation:curleptdexpand}]).
The bottom panels show scatter-plots of the original versus derived values
of the $x-$, $y-$, and $z-$ 
components of $c \grad \times \vecE$.  Note the scatter-plots
for the $x-$ and $y-$ components of 
$c \grad \times \vecE$ are not as tight as the
$z-$ component.  This stems from using finite difference operations for
the Laplacian that aren't strictly compatible with the finite difference
operations we have used to evaluate the curl and divergence operators.
We have also performed similar comparisons for FFT-derived solutions 
(Appendix \ref{app:b}), which can
be applied here, since ANMHD does assume periodic boundary conditions.  
In that case, the corresponding x,y scatter-plots are considerably tighter
than the ones shown here.
All of the grayscale
images are plotted using the same scale.
}
\label{figure:anmhdcurle}
\end{figure}

The top panels of Figure \ref{figure:anmhdcurle} show the curl of
the electric field obtained from time differences in the magnetic
field between adjacent snapshots of the magnetic field evolution.  The
middle panels of the Figure show the reconstructed distribution of
$\grad \times \vecE$ as obtained from equation
(\ref{equation:curleptdexpand}).  
The bottom panels show scatter plots between the known and reconstructed
values of $\grad \times \vecE$.
One can see that the reconstructed
components of $\grad \times \vecE$ show good agreement with the
original values from the MHD code.
%

\begin{figure}
\includegraphics[width=6.5in]{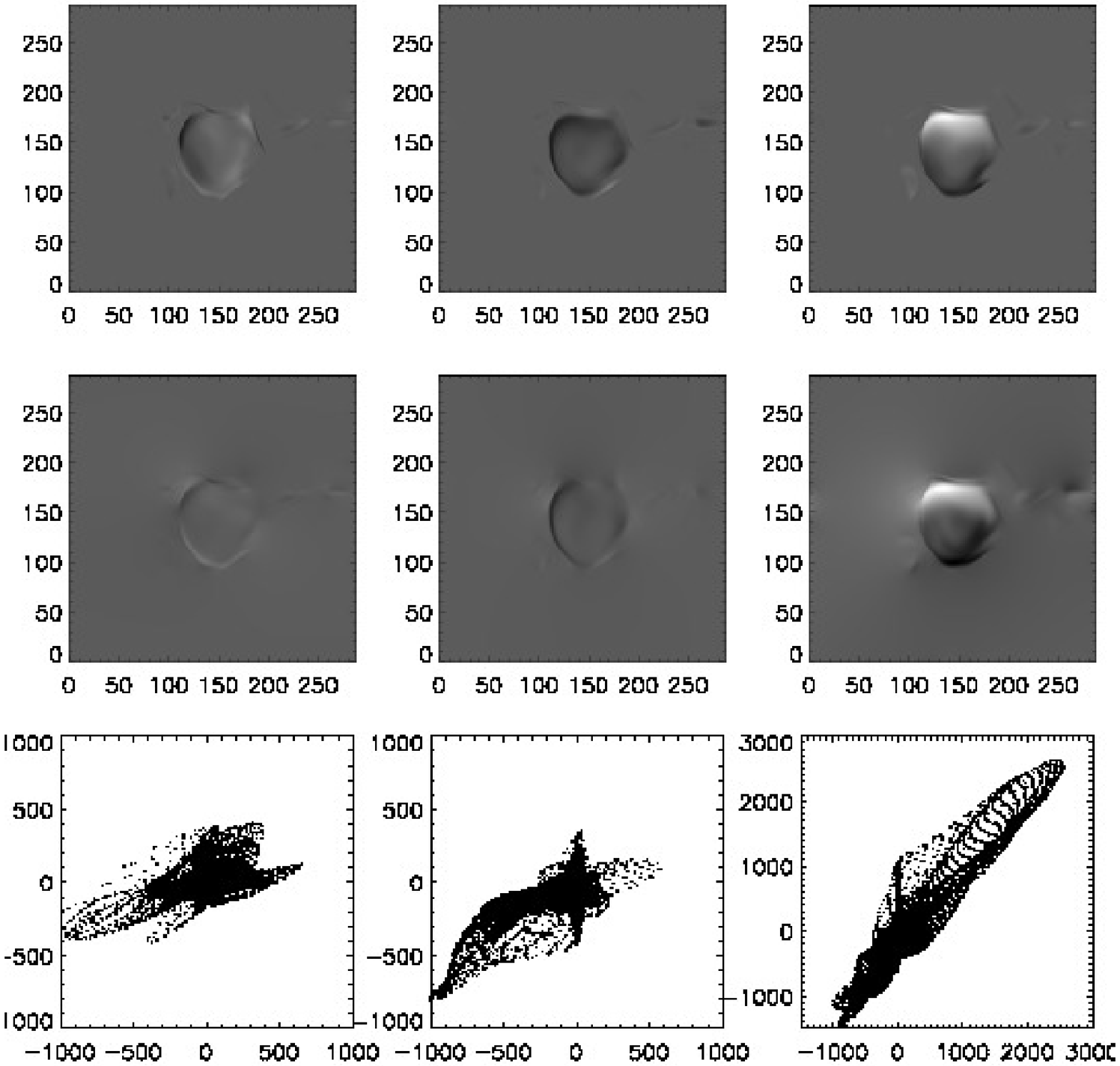}
\caption{
Top panels show the values of the $x-$, $y-$,and $z-$ components of
$c \vecE$ taken from the ANMHD
simulations described in the text.  The middle panels show the same components
of $c \vecE$ determined with the PTD formalism 
(equation [\ref{equation:curleptd}]) assuming $\grad \psi = 0$.
Bottom panels show scatter plots of the $x-$,$y-$,$z-$ 
components of $c \vecE$ derived
from the PTD formalism versus the ANMHD components of $c \vecE$.  Note that
compared to scatter-plots comparing $\grad \times \vecE$ between the
simulation data and the PTD results, the results here show considerable scatter.
All of the grayscale
images are plotted using the same scale.
}
\label{figure:anmhde}
\end{figure}

If one ignores the contribution from $\grad \psi$ in equation
(\ref{equation:eptd}), which is equivalent to assuming that
$\vecE$ is equal to $\vecE^I$, it is straightforward to
compare the reconstructed electric field $\vecE^I$ with that used in
the ANMHD code.  The top panels of Figure \ref{figure:anmhde} show the
actual components of the electric field used in the MHD simulation,
while the middle panels show the electric field components from
equation (\ref{equation:eptd}) approximating $\vecE$ by $\vecE^I$.
The bottom panels show scatter plots of the reconstructed electric field
components as a function of the known values.

The velocity inversion methods tested in \cite{Welsch2007} also included
comparisons between the ANMHD and reconstructed electric field components 
computed by assuming $c \vecE = -\vecv \times \vecB$ (see
their Figures 11 and 12). 
\cite{Welsch2007} showed that the simulation data for $\dot B_z$, computed
by differencing two $B_z$ images in time, matched 
\hbox{$-c \veczhat \cdot (\grad \times \vecE)$}, 
computed using simple second-order finite difference formulae to evaluate
the spatial derivatives applied to the average of the electric fields at the
two adjacent times.
Hence, it
is possible to make a direct comparison between the PTD derived electric
field values reported here
and those from several velocity inversion methods, including
LCT techniques.
The results reported in \cite{Welsch2007} include only ``strong field'' pixel
locations where $| B_z | > 370{\rm G}$, 5\% of the maximum field strength
in this simulation of subsurface magnetic evolution.
Accordingly, to make a direct comparison,
we compute rank-order correlation coefficients between the PTD-derived
electric field components with those from ANMHD, including only the strong
field locations.  The rank-order correlation coefficients 
for the $x-$, $y-$, and $z-$ components of $\vecE$ are $0.76$, $0.87$, and
$0.94$, respectively.  This metric compares favorably with all of the techniques
shown in Figures 11 and 12 of \cite{Welsch2007}, showing better correlation
coefficients for $E_x$ and $E_y$ than all the methods except MEF, and 
comparable correlation coefficients to MEF.  All the methods, including PTD,
show good correlation coefficients for $E_z$.

Despite the good performance of PTD
compared to the velocity inversion methods for strong-field locations, 
a scatterplot comparison between the original and reconstructed electric
fields is poor when {\it all} points are considered (the bottom panels of
Figure \ref{figure:anmhde}).  This contrasts with the
scatterplot comparison between 
actual and reconstructed values of $\grad \times \vecE$ which showed good
agreement.

Why is this?  The problem occurs because of the under-determination of
the electric field.  The PTD formalism guarantees that the electric
field will obey Faraday's law, but no other information about 
additional physical mechanisms
that would determine the electric field has been incorporated.
Any additional electric field
contribution that can be represented by the gradient of a scalar
function is left completely unspecified.  Evidently setting $\grad \psi = 0$ in
the PTD solution (equation \ref{equation:eptd}) is inconsistent with
the electric field as specified in the ANMHD simulation.  Therefore, finding
an equation that better describes $\grad \psi$ is essential for providing
a better performance of the PTD method.

Since $\vecE \simeq - \vecv /c \times \vecB$ in these nearly ideal ANMHD data,
$\vecE$ is perpendicular to $\vecB$.
Considering only the 
$c \vecE^I \equiv - \grad \times \dot \scrB \veczhat - \dot \scrJ \veczhat$ 
contributions from the poloidal and toroidal terms in
equation (\ref{equation:eptd}), and decomposing the vectors into the
directions parallel and perpendicular to $\vecB$, one finds
contributions that are roughly equal in the parallel and perpendicular
directions.  To better reconstruct the actual electric
field, it will be necessary to add a potential electric field that
largely cancels out the components of $\vecE$ that are parallel to
$\vecB$.  The challenge is to derive an equation for $\psi$ from
physical or mathematical principles that does this, while also
yielding a physically reasonable solution for the resulting total
electric field.

\subsection{Deriving an Electric Potential I. -  An Iterative Approach}
\label{subsection:iterative}

Here, we 
describe a technique to determine a potential function that
is consistent with ideal MHD ($\vecE \cdot \vecB = 0$), using a purely ad-hoc
iterative approach.  The total electric field is
\be
\vecE = \vecE^I - \grad \psi\ ,
\label{equation:edef}
\ee
where as before, $\vecE^I$ is the inductive contribution found using
the PTD formalism and $-\grad \psi$ is the potential contribution.  We wish
to define $\psi$ in such a way that the components of $\vecE$ parallel to
$\vecB$ are minimized.

In step 1 of the procedure, we decompose $\grad \psi$ into three orthogonal
directions by writing
\be
\grad \psi = s_1(x,y) \vecbhat + s_2(x,y) \veczhat \times \vecbhat
+ s_3(x,y)\ \vecbhat \times ( \veczhat \times \vecbhat )
\label{equation:s1s2s3}
\ee
Here, $\vecbhat$ is the unit vector pointing in the direction of
$\vecB$.

In step 2, we set $s_1(x,y)$ equal to $\vecE^I \cdot \vecbhat$.  
This ensures that
$- \grad \psi$ acts to cancel the component of $\vecE_I$ parallel to $\vecbhat$.
The function $s_1(x,y)$ will remain invariant during the rest of the iteration
procedure.

In step 3, given the latest guess for $\psi$, we evaluate the functions 
$s_2(x,y)$
and $s_3(x,y)$ by dotting each of the vectors on the right hand side
of equation (\ref{equation:s1s2s3}) with equation (\ref{equation:s1s2s3}), 
itself, yielding
\be
s_2(x,y)={\veczhat \cdot (\vecb_h \times \grad_h \psi )  \over b_h^2}\ ,
\label{equation:s2def}
\ee
and
\be
s_3(x,y) = {\partial \psi \over \partial z} - {(\grad_h \psi \cdot \vecb_h)
b_z \over b_h^2}\ .
\label{equation:s3def}
\ee
Here, $b_z$ and $b_h$ represent the amplitudes of $\vecbhat$ in the vertical
and horizontal directions, respectively, and $\vecb_h$ represents only the
horizontal components of $\vecbhat$.  For the initial guess during step
3, the functions $s_2(x,y)$ and $s_3(x,y)$
are set to zero.

In step 4, the horizontal divergence of $\grad_h \psi$ 
is taken, using the current
guesses for $s_2(x,y)$ and $s_3(x,y)$.  From equation (\ref{equation:s1s2s3})
this results in the following two-dimensional Poisson equation for $\psi$:
\be
\nabla_h^2 \psi = \grad_h \cdot ( s_1(x,y)\vecb_h + s_2(x,y)(\veczhat \times
\vecbhat ) - s_3(x,y) b_z \vecb_h )
\label{equation:poisson_iterative}
\ee
The values for $\psi$ are updated by solving this Poisson equation.

In step 5, the vertical gradient of $\psi$ is updated 
by evaluating the z-component of equation 
(\ref{equation:s1s2s3}) employing the last guess for
$s_3(x,y)$ (recall that $s_1(x,y)$ does not change between iterations):
\be
{\partial \psi \over \partial z} = s_1(x,y) b_z + s_3(x,y) b_h^2\ .
\label{ref:dpsidz_iterative}
\ee

Step 6 consists of evaluating an error term
\be
\epsilon = max { |( \vecE_I - \grad \psi ) \cdot \vecbhat | \over
| ( \vecE_I - \grad \psi ) | } \ .
\ee
If $\epsilon$ is sufficiently small, then the iteration sequence can be
terminated; otherwise steps 3-6 are repeated until the sequence converges
to the desired error criterion.

\begin{figure}
\includegraphics[width=6.5in]{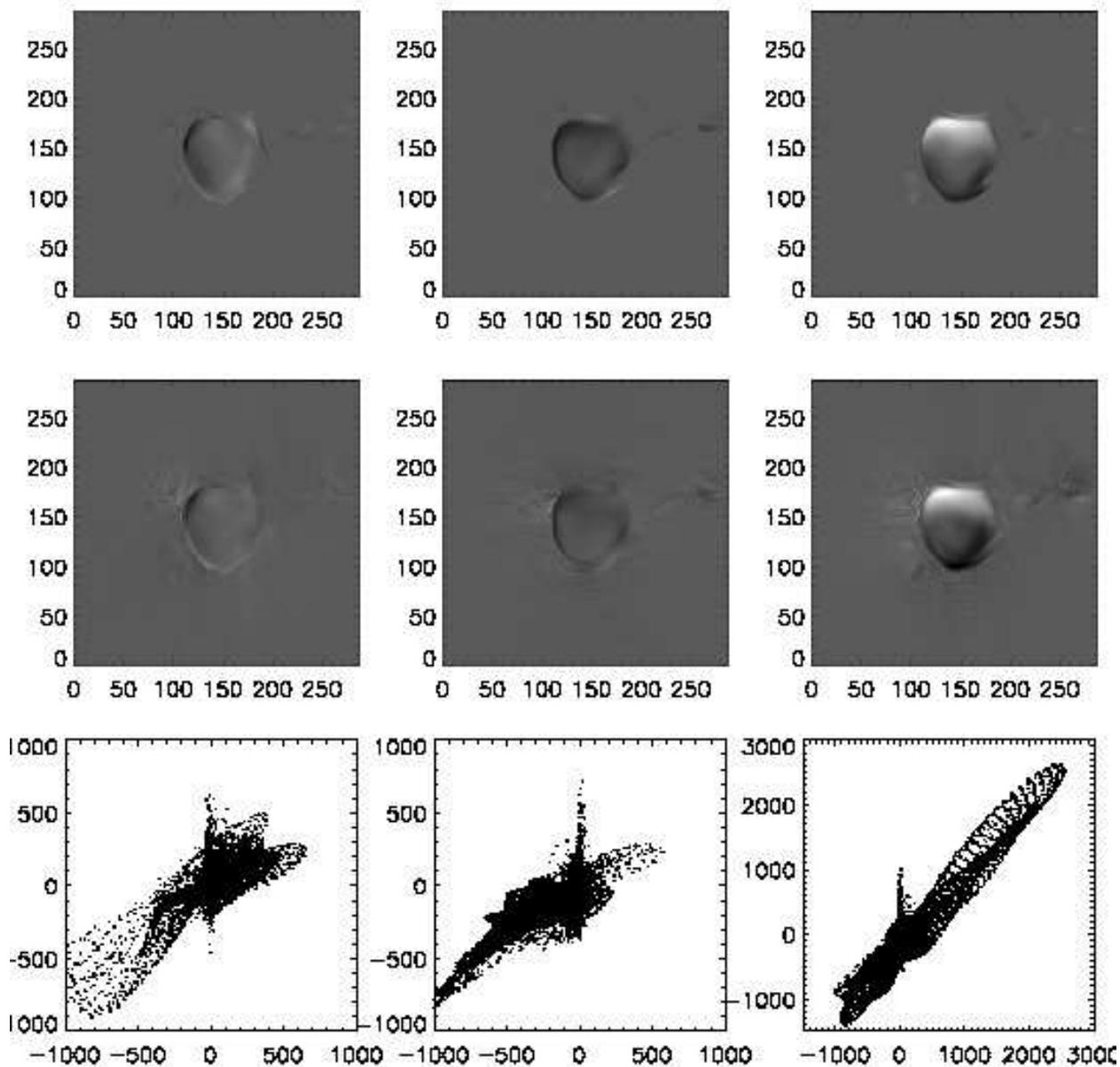}
\caption{
Top panels show the values of the $x-$, $y-$, and $z-$ components of
$c \vecE$ taken from the ANMHD
simulations described in the text.  The middle panels show the same components
of $c \vecE$ determined with the PTD formalism 
(equation [\ref{equation:curleptd}]) adding $- \grad \psi$ as determined
with the ``iterative'' technique of \S \ref{subsection:iterative}.
Bottom panels show scatter plots of the $x-$, $y-$,and $z-$ 
components of $c \vecE$ derived from the PTD plus potential field solutions 
versus the ANMHD components of $c \vecE$.  Note that
compared to scatter-plots comparing $c \vecE$ from PTD solutions without
the potential contribution (Figure [\ref{figure:anmhde}]), 
the results here show less scatter.
All of the grayscale
images are plotted using the same scale.
}
\label{figure:iterative}
\end{figure}

Results of this iteration sequence applied to the PTD solutions, and compared
to the true electric field results from the ANMHD simulation are shown in
Figure \ref{figure:iterative}.  To solve the Poisson equation 
(\ref{equation:poisson_iterative}) in this example, we
take advantage of the known periodic nature of the ANMHD solutions, and use
FFT techniques to solve the equation for $\psi$.  

From Figure \ref{figure:iterative}, one can see that some
of the artificial features in the PTD recovered solutions have been improved
by applying the iteration scheme described here, such as the false bright
halo seen in the PTD solution for $E_x$ on the upper left side.  Further,
scatterplots of $E_x$ and $E_y$ comparing results between simulation values
and those with PTD only, and those with PTD plus the iteration scheme, show
clear improvement by adding in the contributions from this potential function.
However, the technique is by no means perfect.  After applying the 
$\grad \psi$ correction derived with the iteration technique, other ``shadow''
artifacts, seen as faint vertical stripes below the emerging magnetic field,
are visible in $e.g.$ the derived map of $E_x$.  Computing the rank-order
correlation coefficients between the ANMHD and iteration method electric fields
in the strong magnetic field regions, as we did
for the PTD solutions, results in values of $0.75$, $0.82$, and $0.95$ for
the $x-$, $y-$ and $z-$ components of $\vecE$.  These values are not
significantly different than those for the PTD case described earlier.

In summary, this iteration scheme, or perhaps similar schemes based on
related ideas, may provide a useful approach for 
a deriving potential electric field contribution which,
when added to the PTD solutions, is consistent with ideal 
MHD.  But we must also caution that the
solutions derived via this method are not unique 
since
the condition $\vecE \cdot \vecB = 0$ does not fully constrain
the potential function $\psi$.  Once a solution has been obtained via this
method, we can add on any additional solution $\psi '$ which obeys the
constraint $\grad \psi ' \cdot \vecB = 0$, without affecting the induction
equation or the condition $\vecE \cdot \vecB = 0$.

It is also not clear whether
this iteration technique will converge for all cases, or 
whether the derived solutions are mathematically well posed apart from the
uniqueness issue already raised.  Clearly this area needs further study.

\subsection{Deriving an Electric Potential II. -  A Variational Approach}
\label{subsection:variational}

The dynamics of the solar plasma is determined by the
largest forces, which in regions of strong magnetic fields will involve
Lorentz forces, acting in conjunction with gravity, pressure
gradients, and inertial terms.  
To the extent that the electric field is 
dominated by the ideal $- \vecv / c \times \vecB$ term, 
then it is necessary to know the forces acting to determine the velocity
field in the photosphere to determine the full electric field 
and thus completely specify $\psi$.

In \S \ref{subsection:nopsi}, we demonstrated that vector magnetograms alone 
contain only partial
information about the plasma dynamics -- there simply isn't enough information
in the magnetic field data alone to uniquely specify $\vecv$ or
$\vecE$.  Additional information must be obtained either from other 
measurements or by using some other constraint.

One approach for deriving a constraint 
equation for $\psi$ is to use a variational
principle.  For example, one could adjust $\psi$ such that the 
electromagnetic field energy density, \hbox{$(E^2+B^2)/(8 \pi)$,} 
integrated over the
magnetogram, is minimized.  Since $\vecB$ itself has already been determined
by the measurements, this is tantamount to finding $\psi$ such that the area
integral of $E^2$ is minimized.  The motivation for this approach
is to reduce or eliminate the unphysical electric field ``halos'' seen
in regions of negligible magnetic fields strength in Figure
\ref{figure:anmhde}.  An alternative equation for $\psi$ can be
derived by minimizing $| c \vecE \times \vecB / B^2 |^2$ integrated over the
magnetogram, which is equivalent to minimizing the kinetic energy of flows
in the photosphere, if one assumes $\vecE = - \vecv / c \times \vecB$.  This
is essentially the approach used by \cite{Longcope2004} in his MEF technique
for deriving flows from magnetograms using only the vertical component of
the induction equation.

Here we derive an equation for $\psi$ which is sufficiently general that both
of these cases can be included using the same formalism.  We
minimize the functional
\be
L = \int \, dx \, dy \,  W^2(x,y) [
(E^I_x - \partial \psi / \partial x)^{2} +
(E^I_y - \partial \psi / \partial y)^{2} +
(E^I_z - \partial \psi / \partial z)^{2} ] ~,
\label{equation:w2min}
\ee
where $W^2(x,y)$ is an arbitrary weighting function, and $E_x^I$, $E_y^I$,
and $E_z^I$ are the three components of $\vecE^I$ as it defined in
equation (\ref{equation:eptd}).

In the photosphere, we believe the electric field $\vecE^I - \grad \psi$
is dominated by the ideal term $-\vecv / c \times \vecB$, but with a possible
additional contribution $\vecR$, which can represent resistive or any 
other non-ideal
electric field terms.  We assume that $\vecR$ is either a known function, 
is determined by observation, or is specified by the user as an Ansatz. Then
\be
\vecE = -{\vecv \over c} \times \vecB + \vecR = \vecE^I -\grad \psi\ .
\label{equation:billwants}
\ee
By dotting $\vecE^I - \grad \psi$ with $\vecB$, one finds
\be
( \vecE^I - \grad \psi ) \cdot \vecB = \vecR \cdot \vecB .
\label{equation:Rdef}
\ee
This equation provides an additional constraint on the potential $\psi$,
allowing us to eliminate $\partial \psi / \partial z$ in favor of
$\grad_h \psi$:
\be
B_z {\partial \psi \over \partial z} = {\bf B} \cdot \vecE^I
- {\bf B}_h \cdot {\grad}_h \psi - \vecR \cdot \vecB .
\label{equation:idealbhat}
\ee

Note that the functional minimized in equation 
(\ref{equation:w2min}) depends on $\psi$ through its
dependence on the total electric field $\vecE^I-\grad \psi$.  In particular,
the z-component of $\vecE$ that appears in equation (\ref{equation:w2min})
depends on $E_z^I - \partial \psi / \partial z$.  But from the above
constraint equation (\ref{equation:idealbhat}) we can see that
\be
B_z ( E_z^I - {\partial \psi \over \partial z} ) =
\vecR \cdot \vecB - {\bf B}_h \cdot (\vecE_h^I - \grad_h \psi ) , 
\label{equation:jgone}
\ee
showing that $E_z^I = -\dot \scrJ$ cancels out of the variational
equation.  This result shows that a solution of the variational
problem for $\vecE^I - \grad \psi$ will be independent of the solutions for
$\dot \scrJ$ and $\partial \dot \scrB / \partial z$ as determined from equations
(\ref{equation:poissonjz}-\ref{equation:poissondivbh}) 
and boundary conditions (\ref{equation:bdrydbdz}-\ref{equation:bdryj}).  
Thus the fact that 
$\partial \dot \scrB / \partial z$ and $\dot \scrJ$
do not have unique solutions does not
affect the uniqueness of the solutions for the electric field itself:
In this approach,
changes in $\dot \scrJ$ are compensated by changes 
in $\partial \psi / \partial z$ such that $E_z$ is unchanged.

Performing the Euler-Lagrange minimization of equation (\ref{equation:w2min}) 
results in a second-order, two-dimensional elliptic partial
differential equation for $\psi$,
\be  
0 = \grad_h \cdot \left \{ W^2(x,y) \left (
(\vecE^I_h - \grad_h \psi) + {\bf B}_h
\frac{ \left [{\bf B}_h \cdot (\vecE^I_h - \grad_h \psi) -\vecR \cdot \vecB 
\right ] }
{B_z^2} \right ) \right \}
\label{equation:eulergeneral}
\ee
Equation (\ref{equation:eulergeneral}) involves a combination of both the
inductive electric field from equation (\ref{equation:eptd}) and the
potential contribution from $- \grad_h \psi$, assuming that $\vecE^I$ is given.
Solving this equation for $\psi$ given $\vecE_h^I$ 
is essentially
the approach taken by \cite{Longcope2004} in the development of MEF.

Alternatively, equation (\ref{equation:eulergeneral})
can be viewed as a single equation for the sum of
both the inductive and potential contributions, 
to be determined simultaneously.  If the equation for the total
field can be solved, 
then the potential term can be found afterward
by subtracting the PTD
solution (equation [\ref{equation:eptd}]) from the
solution for the total electric field.

Writing the total electric field $\vecE$ as $\vecE^I-\grad \psi$,
or $E_z=E_z^I-\partial \psi / \partial z$, and $\vecE_h=\vecE_h^I-\grad_h \psi$,
and noting that equation (\ref{equation:jgone}) relates $E_z$ to $\vecE_h$,
equation (\ref{equation:eulergeneral}) can be re-written as
\bea
\grad_h \cdot ( W^2/B_z ) \left ( \vecE_h B_z - E_z\vecB_h \right ) &=&
\nn
\\ 
- \grad_h \cdot ( W^2/B_z ) \left ( \vecE \times \vecB \right ) 
\times \veczhat &=&
\nn
\\
-\veczhat \cdot \grad_h \times
( W^2/B_z )  \left ( \vecE \times \vecB \right )_h &=& 0 .
\label{equation:xmefnocurl}
\eea
The variational approach thus leads to a local condition on the quantity
$( W^2/B_z ) ( \vecE \times \vecB )_h$, namely that it is curl-free,
and thus can be represented as 
the gradient of a two-dimensional scalar function.
Therefore, we write
\be
( W^2/B_z ) ( c \vecE \times \vecB )_h = - \grad_h \chi .
\label{equation:chidef}
\ee

We wish to derive an equation for $\chi$, starting from 
equation (\ref{equation:chidef}) and involving only known quantities from
the vector magnetic field or its time derivatives.  
The details of the
derivation are shown in Appendix \ref{app:cd}.  The result is
\be
{- \partial B_z \over \partial t} + \grad_h \cdot 
\left( {c \vecR \cdot \vecB \over B^2}\ \veczhat
\times \vecB_h \right) + \grad_h \cdot \left( {1 \over W^2 B^2}  
( \vecB_h \cdot \grad_h \chi ) \vecB_h \right) = - \grad_h \cdot 
\left( {B_z^2 \over W^2 B^2 } \grad_h \chi \right)  . 
\label{equation:chidivnoapp}
\ee
Eventually, we will assume $\vecR=0$, but for now we retain it in
our formalism so that non-ideal effects can be included.

If one either knows $\vecR$ or sets $\vecR \cdot \vecB$ to $0$,
equation (\ref{equation:chidivnoapp}) 
is a two-dimensional linear elliptic partial
differential equation for $\chi$, with coefficients that depend on 
the magnetic field components or its time derivatives.  
Thus we can regard the solution
for $\chi$ as well-defined.  

In Appendix \ref{app:cd} we determined how to find $E_z$ from $\chi$ 
via equation (\ref{equation:ezoverbz}).  
To find $\vecE_h$, one can take the cross-product
of $\veczhat$ with equation (\ref{equation:chidefoverw2}) and use equation
(\ref{equation:ezoverbz}) to find
\be
c \vecE_h = c {\vecR \cdot \vecB \over B^2} \vecB_h - {B_z^2 \over W^2 B^2}\ 
\veczhat \times \grad_h \chi - {1 \over W^2 B^2}\ ( \vecB_h \cdot \grad_h \chi )
\ \veczhat \times \vecB_h .
\label{equation:ehfromchi}
\ee
The Poynting flux $\vecS \equiv ( c / (4 \pi )) \ \vecE \times \vecB$ has
components that point along the gradient of $\chi$
in the horizontal direction as the definition (\ref{equation:chidef}) shows.
However, we can determine the Poynting flux in the vertical direction
as well by noting that $\vecS \cdot \vecB = 0$, or 
$B_z S_z = - \vecB_h \cdot \vecS_h$.  Thus we find
\be
\vecS_h = -{B_z \over 4 \pi W^2} \grad_h \chi ,
\label{equation:sh}
\ee
and
\be
S_z = {\grad_h \chi \cdot \vecB_h \over 4 \pi W^2} .
\label{equation:sz}
\ee

Assuming that the non-ideal electric field term $\vecR$ is zero or negligible 
compared to
the ideal contribution, equation (\ref{equation:chidivnoapp}) is greatly 
simplified for the two special cases of $W^2=1/B^2$ (minimum kinetic energy)
and $W^2=1$ (minimum electric field energy).  In those cases, equation
(\ref{equation:chidivnoapp}) becomes 
\be
(W^2 B^2 = 1:)\ \ \ 
{- \partial B_z \over \partial t} 
+ \grad_h \cdot \left(  
( \vecB_h \cdot \grad_h \chi ) \vecB_h \right) = - \grad_h \cdot 
\left( B_z^2 \grad_h \chi \right)  . 
\label{equation:chiformekf}
\ee
and
\be
(W^2 = 1:)\ \ \ \ 
{- \partial B_z \over \partial t} 
+ \grad_h \cdot \left(  
( \vecb_h \cdot \grad_h \chi ) \vecb_h \right) = - \grad_h \cdot 
\left( b_z^2 \grad_h \chi \right)  . 
\label{equation:chiformeef}
\ee
In equation (\ref{equation:chiformeef}) $b_z$ and $\vecb_h$ represent, 
respectively, the vertical and horizontal components of the unit vector 
$\vecbhat$ pointing in the direction of the magnetic field.  

Once the variational equation for $\chi$
has been solved, how does one relate the
total electric field to the PTD solutions and the potential field contribution?
Since the total electric field
$\vecE = \vecE^I-\grad \psi$, we can subtract the $\vecE^I$ contribution
of equation 
(\ref{equation:eptd}) 
from equations
(\ref{equation:ehfromchi}) and (\ref{equation:ezoverbz}) 
to derive these expressions for the electric field from $\grad \psi$:
\be
-c \grad_h \psi = 
c {\vecR \cdot \vecB \over B^2} \vecB_h - {B_z^2 \over W^2 B^2}\ 
\veczhat \times \grad_h \chi - {1 \over W^2 B^2}\ ( \vecB_h \cdot \grad_h \chi )
\ \veczhat \times \vecB_h + \grad_h \times \dot \scrB \veczhat
\label{equation:gradhpsi}
\ee
and
\be
-c {\partial \psi \over \partial z} =  B_z {c \vecR \cdot \vecB \over B^2} - 
{B_z \grad_h \chi  \cdot ( {\veczhat \times \vecB_h}) \over W^2 B^2 } + 
\dot \scrJ .
\label{equation:dpsidz}
\ee


\begin{figure}
\includegraphics[width=6.5in]{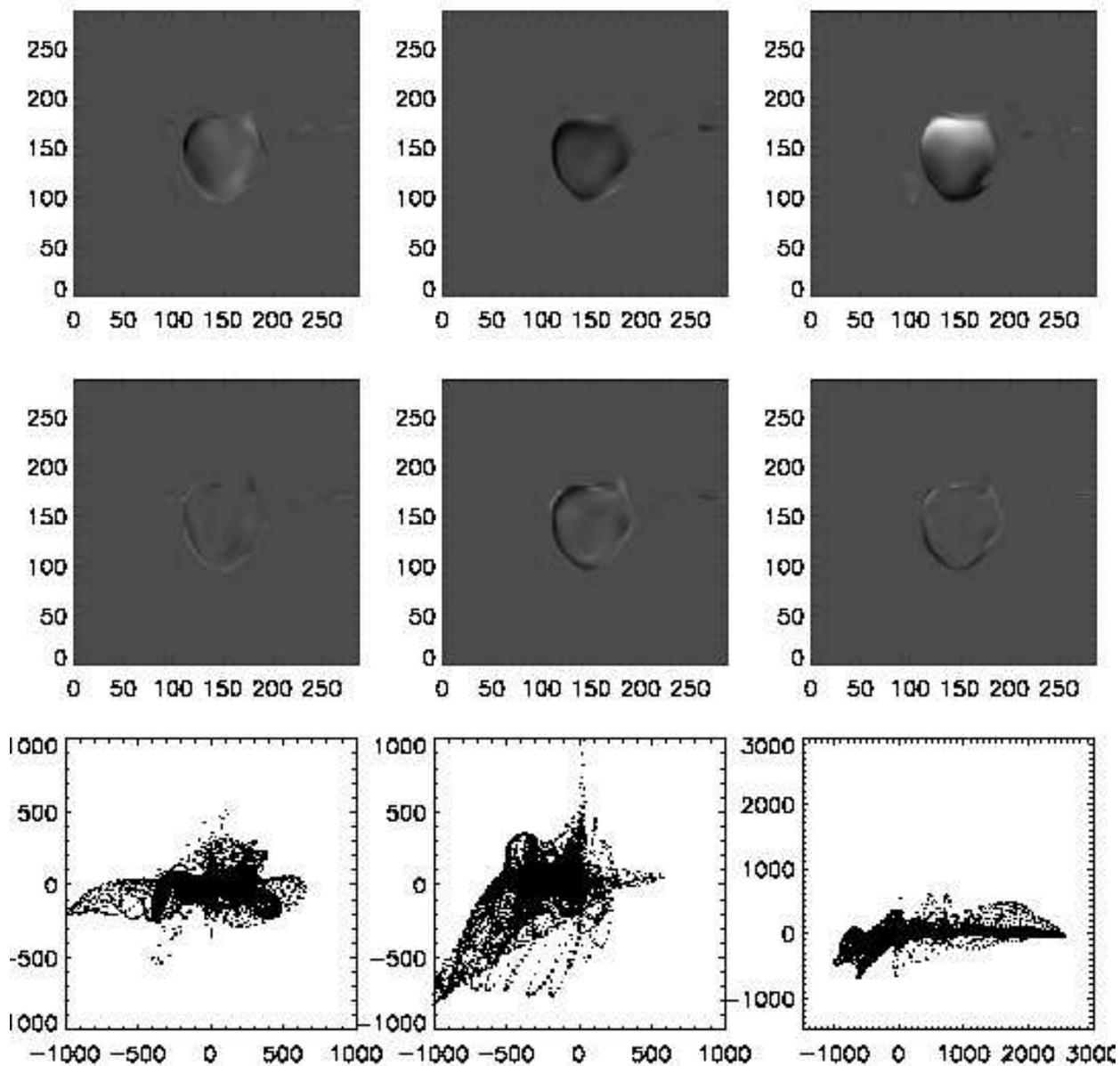}
\caption{
Top panels show the values of the $x-$, $y-$, and $z-$ components of
$c \vecE$ taken from the ANMHD
simulations described in the text.  The middle panels show the same components
of $\vecE$ determined with the variational formalism 
(equation [\ref{equation:chiformeef}], plus equations [\ref{equation:ezoverbz}]
and [\ref{equation:ehfromchi}], $i.e.$ minimizing $E^2$).
The bottom panels show scatter-plots of the original versus derived values
of the $x-$, $y-$, and $z-$ components of $c \vecE$.  
All of the grayscale
images are plotted using the same scale.
}
\label{figure:fig_var}
\end{figure}

The variational equation for $\chi$ 
incorporates the vertical component of the induction equation (the part
that depends on ${\partial B_z}/{\partial t}$),
but does not depend at all on ${\partial \vecB_h}/{\partial t}$.
This means, for the variational solutions, that any observed time 
behavior for $\vecB_h$ can be specified independently of the time 
behavior for $B_z$.

To illustrate the variational technique, we minimize $E^2$ integrated
over the magnetogram, ($W^2=1$) again using the ANMHD simulation described in 
\S \ref{subsection:nopsi}. The ``observed'' map of 
$\partial B_z / \partial t$ is used as input, and we solved equation 
(\ref{equation:chiformeef}) for $\chi$, using Neumann boundary conditions 
(assuming zero horizontal
Poynting flux entering at the horizontal boundaries).  
This is a very good approximation for all but a few short segments of this
synthetic magnetogram boundary.
Since equation (\ref{equation:chiformeef}) shows that
$\chi = 0$ is a solution in regions of insignificant $\dot B_z$,
corresponding to the low field strength regions of the domain,
the solution for $\chi$ is set to zero for
magnetic fields strengths below a threshold value.  Once a solution for $\chi$
is found, we verified that the horizontal components of $\vecE$
found from equation (\ref{equation:ehfromchi}) obeyed the induction equation,
$i.e.$, $c \grad_h \times \vecE_h = - \partial B_z / \partial t$.  The resulting
three components of $\vecE$ are shown as the middle panels
in Figure \ref{figure:fig_var}.  

The results show a generally poor agreement
with $\vecE$ from the ANMHD simulation.  Thus, at least in this case,
the variational approach of minimizing $E^2$ does not do a good job of 
reproducing the actual electric field.  The PTD solution by itself ($i.e.$
assuming that $\vecE = \vecE^I$) shows better agreement with the simulation
data, even though it produces spurious components of $\vecE$ parallel
to $\vecB$ (which the ANMHD simulations did not have).  Computing the rank-order
correlation coefficients between the ANMHD and variational electric fields
in the strong magnetic field regions results in values of $0.33$, $0.50$, and
$0.50$ for the $x-$, $y-$, and $z-$ components of $\vecE$, also indicating
a poor agreement with the ANMHD electric field results.

In spite of the poor comparison with the ANMHD results,
the variational method did what it was designed to do: find an electric
field that obeyed the induction equation, yet do this with minimum
amplitude.  The ``halos'' shown in the PTD solutions of Figure
\ref{figure:anmhde}, for example, have been eliminated.
Although the resulting electric field was not consistent with
the simulation data, the variational method does yield a physically 
reasonable result, and stays zero in regions where one expects
to find $\vecE$ near zero.
These results motivate future exploration of other choices for $W^2$, to
determine whether other choices result in better fits of the variational
results to the MHD simulation data.


%

\section{An Example: NOAA AR 8210}
\label{section:8210}

To illustrate the ideas described in the previous sections with
an example using a real
sequence of vector magnetograms, we apply
these techniques to a pair of vector magnetograms of NOAA Active Region 8210 
taken with the University
of Hawaii's Imaging Vector Magnetograph instrument (IVM) at Mees 
Solar Observatory on Haleakala \citep{Mickey1996}.
These observations have already been described in detail
in \cite{Welsch2004}.  We chose this observational test case because the data
is well known, and we can therefore
forego a detailed discussion of the data and its analysis.

The pair of vector magnetograms, separated by a period of roughly 4 hours,
were first cross-correlated and shifted 
to remove a mean shift due to solar rotation, and then averaged to
define a mean vector magnetic field, and differenced to approximate
a partial time derivative of each magnetic field component. 

We first apply the PTD formalism for the average vector magnetic field to
derive the three fields $\scrB$, $\partial \scrB / \partial z$, and $\scrJ$.
In \S \ref{subsection:boundaries} we pointed out 
that the solutions are degenerate,
in that solutions to the homogeneous Cauchy-Riemann equations can be added
to the solutions for $\partial \scrB / \partial z$ and $\scrJ$ without
affecting the derived values of $\vecB_h$.  The degeneracy can be removed by
choosing to set either $\partial \scrJ / \partial s\ = 0$, or 
$\partial / \partial s\ ( \partial \scrB / \partial z ) = 0$ (but not both)
when applying boundary conditions (\ref{equation:bdrydbdz}) and 
(\ref{equation:bdryj}) to the solutions of the Poisson equations for these
two functions.  
We show in Figure \ref{figure:cauchy} how the functions 
$\partial \scrB / \partial z$ and $\scrJ$ differ depending on which parallel
derivative is set to zero along the magnetogram boundary.  Neither choice
affects the derived values of $\vecB_h$, but there is a slight advantage
to choosing to set $\partial \scrJ / \partial s = 0$:  in that case, the
observed values of $B_n$ are due entirely to 
$\partial / \partial n (\partial \scrB / \partial z )$, and therefore
correspond to the potential-field solution that matches $B_n$ at the
boundaries (Appendix \ref{app:a}).  That means that any contribution to
the horizontal magnetic field from currents can be attributed entirely to
the contribution from $\scrJ$.  

\begin{figure}
\includegraphics[width=6.0in]{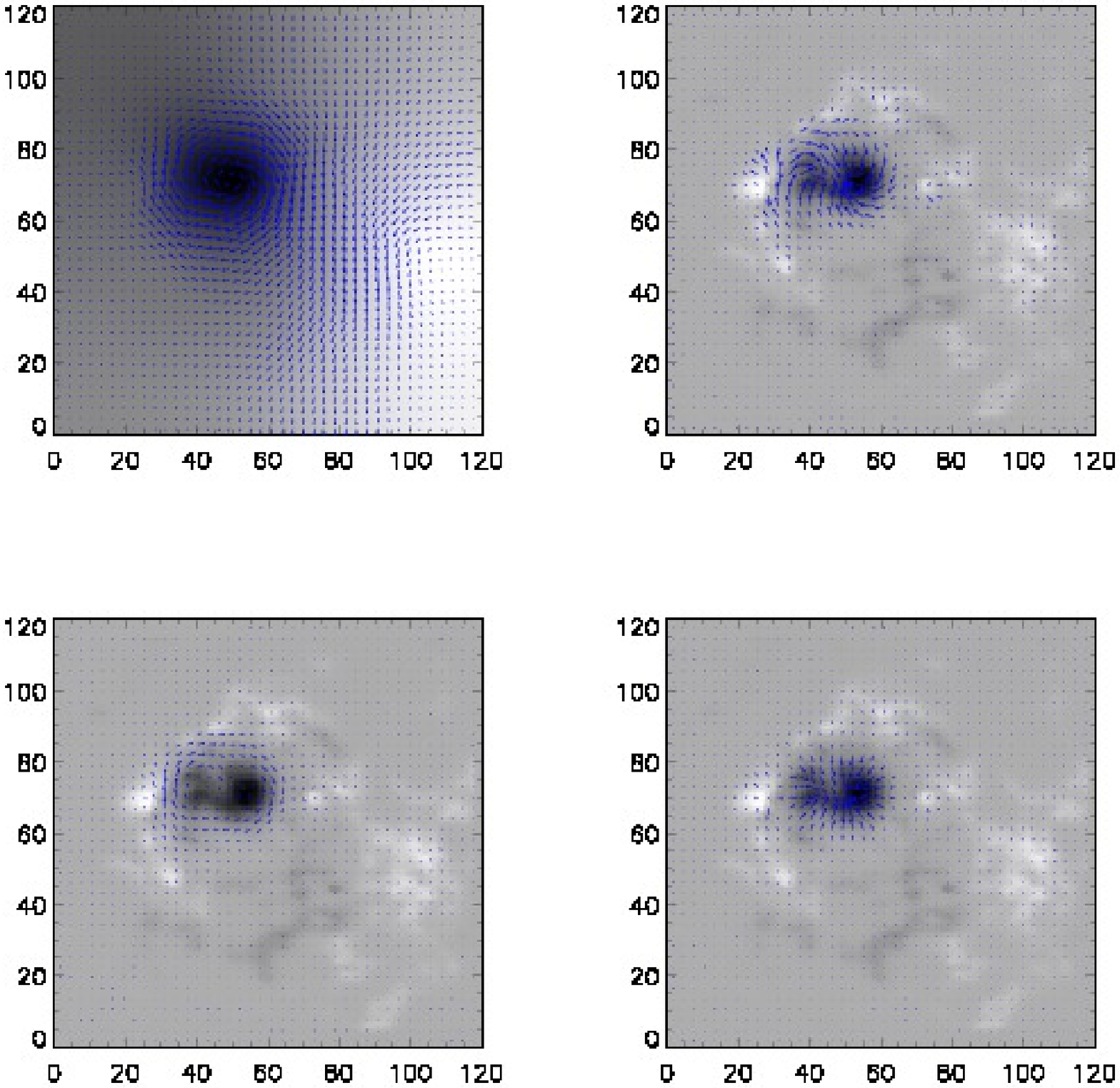}
\caption{
The upper left panel shows $\scrB$ as the background image, while 
$\vecA_P = \grad_h \times \scrB \veczhat$,
the vector potential for the potential magnetic field with the same
$B_z$ as the vector magnetogram, is shown as the arrows.  
The arrows in the upper
right panel show $\vecB_h$, which can be decomposed into the
two contributions shown in the lower two panels.  
The lower left panel, also with $B_z$ as the background image, shows the
contribution to $B_h$ solely from $\grad_h \times \scrJ \veczhat$, which shows
the contributions from non-zero values of $J_z$.  
The lower right panel
shows the vertical magnetic field $B_z$ as the background image, and the
horizontal components of the potential magnetic field 
$\vecB_h^P = \grad_h \partial \scrB / \partial z$ 
(equation [\ref{equation:bhpot}]) as the arrows.  
All of the horizontal magnetic
field vectors are drawn at the same scale.
}
\label{figure:bpot}
\end{figure}

These points are illustrated in Figure
\ref{figure:bpot}, which shows the spatial distribution of $\scrB$,
$\vecA_P = \grad_h \times \scrB \veczhat$, $B_z$, $\vecB_h$,
and the contributions of both the potential-field and current sources to
$\vecB_h$.  To remove the Cauchy-Riemann degeneracy, it was 
assumed that $\partial \scrJ / \partial s = 0$ along the magnetogram boundary,
coinciding with the choice of the top two panels of Figure \ref{figure:cauchy}.
A homogeneous
Neumann boundary condition is used to compute 
$\scrB$:  $\partial \scrB / \partial n = 0$ along the magnetogram boundaries.

Considering next the time evolution of the magnetic field, we first apply the
PTD solution to the measured values $\partial \vecB / \partial t$.  To solve
equation (\ref{equation:poissonbz}) for $\dot \scrB$, 
we have to assume a boundary
condition for $\vecE$ at the edges of the magnetogram.  Generally, this is
not known, but if one is sufficiently lucky to have the magnetogram
boundary located entirely in a region with zero or small magnetic fields,
it is reasonable to assume that $\vecE=0$ along the magnetogram boundary.
If one is solving only the equation for $\dot \scrB$ 
(and not solving for $\psi$ simultaneously), one can
only require that one component of $\vecE_h$ vanish along the boundary.  
We believe it is more physical to set the component of $\vecE_h$ parallel to the
boundary ($E_s$) to zero at the boundary, which means setting the normal
derivative of $\dot \scrB$ to zero there (homogenous Neumann 
boundary conditions), motivated by the discussion 
in \S \ref{subsection:boundaries}.

Our data for AR8210 includes some regions that have significant
magnetic field changes along short sections of
the magnetogram boundary.  Since we are merely
trying to demonstrate how to use these techniques with real data, our approach 
here is to
set $\partial \vecB / \partial t = 0$ in a narrow strip of three zones just
inside the magnetogram boundary, and assume we can then set the parallel
component of $\vecE_h$ ($E_s$) in this slightly altered test case to zero.  
This boundary condition is equivalent to
having a zero derivative of $\dot \scrB$ in the direction normal to the 
boundary (Neumann boundary conditions).  Once this assumption is made, it is
also advisable to use the same Neumann boundary conditions for $\scrB$, so that
the time evolution of $\vecA$ and $\vecA_P$ are consistent with $\vecE$
computed at the boundaries.

If one is using only the PTD solutions, it is probably also a good idea to
set a physically reasonable boundary condition for 
$\dot \scrJ$ at the edge of the magnetogram.
For reasons similar to those described above, it seems reasonable to assume
that $E_z=0$ along magnetogram boundaries that lie in regions of weak average
field.  Since $c E_z = - \dot \scrJ$ in the PTD formalism, this can be
achieved by choosing to set $\partial \dot \scrJ / \partial s \  = 0$ when
applying the boundary conditions 
(\ref{equation:bdrydbdz}-\ref{equation:bdryj}) to 
solving the Poisson equations 
(\ref{equation:poissonbz}-\ref{equation:poissondivbh}).  
Making this assumption removes the Cauchy-Riemann degeneracy for the 
$\dot \scrJ$ and $\partial {\dot \scrB} / \partial z$ solutions.  This
has the added consequence of allowing one to interpret the contributions
of $\partial {\dot \scrB} / \partial z$ to $\partial \vecB_h / \partial t$
as being changes to the potential-field part of the solution.

\begin{figure}
\includegraphics[width=6.0in]{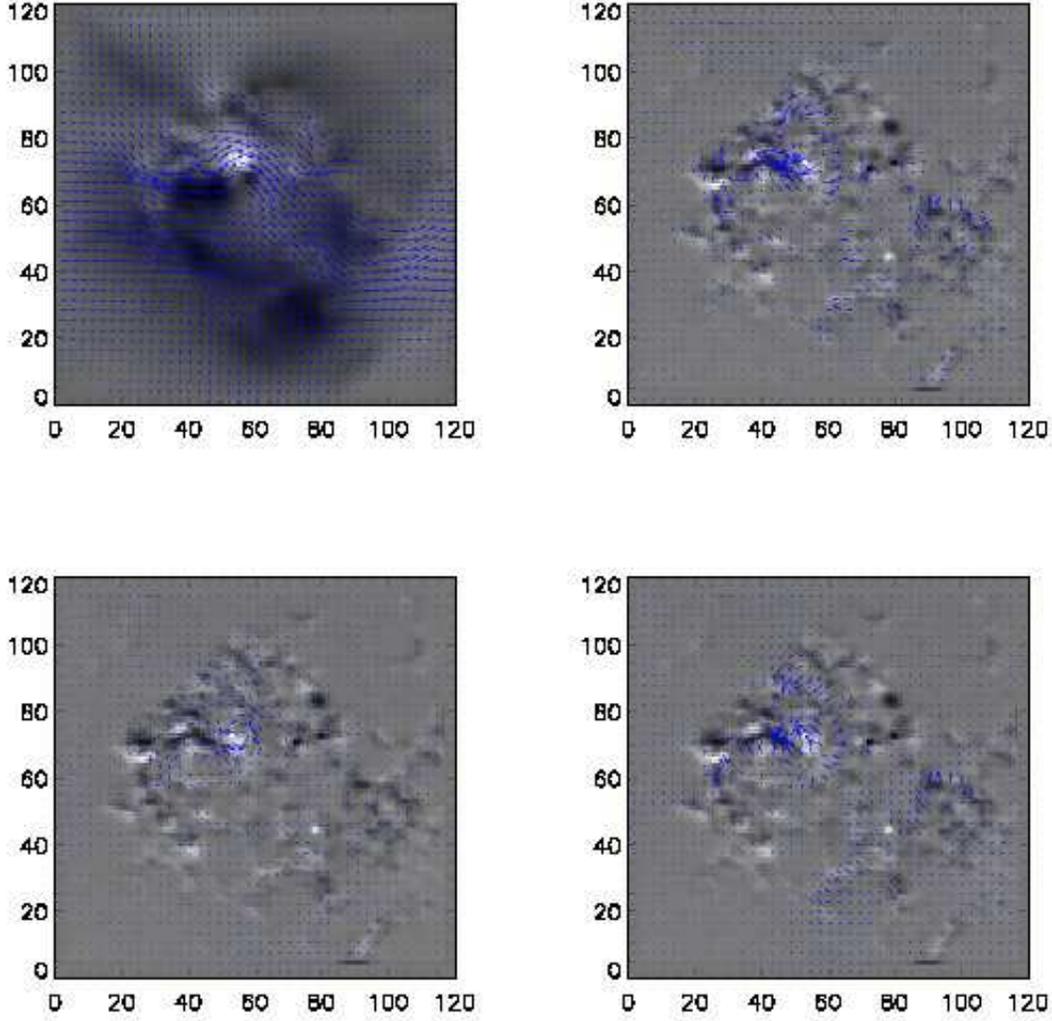}
\caption{
The upper left panel shows $c E_z^I = -\dot \scrJ$ 
as the background image, while 
$c \vecE_h^I = - \grad_h \times \dot \scrB \veczhat$,
the horizontal electric field vector responsible for the time evolution of
$B_z$ from the vector magnetogram difference, is shown as the arrows.  
The arrows in the upper
right panel show 
$\partial \vecB_h / \partial t$, with $\partial B_z / \partial t$ as the
background image.  
The lower left panel, also with $\partial B_z / \partial t$ 
as the background image, shows the
contribution to $\partial \vecB_h / \partial t$ 
solely from $- \grad_h \times \dot \scrJ \veczhat$, which shows
the contributions from non-zero values of $\partial J_z / \partial t$.  
The lower right panel
shows $\partial B_z / \partial t$ as the background image, and 
the time evolution of the
horizontal components of the potential magnetic field,
$\partial \vecB_h^P / \partial t = - \grad_h \partial \dot \scrB / \partial z$ 
as the arrows.  The vectors in the three panels showing changes to $\vecB_h$ are
all drawn at the same scale.
}
\label{figure:dynamic}
\end{figure}

Figure \ref{figure:dynamic} shows the resulting PTD solutions for $\vecE^I$,
and for $\grad \times \vecE^I$.  Also shown are the decompositions of
the curl into the evolution of the potential-field, and those driven by the 
observed evolution in $J_z$.

\begin{figure}
\includegraphics[width=6.5in]{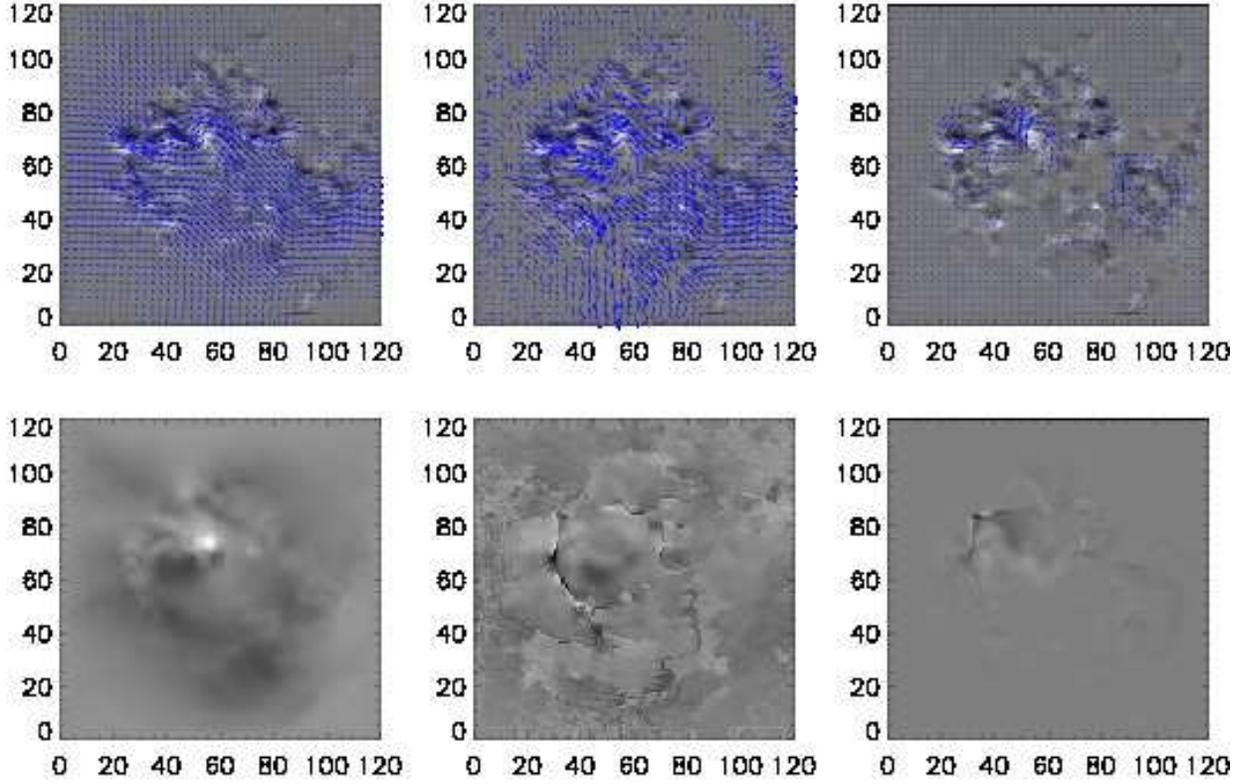}
\caption{
The top three panels show the derived horizontal electric field using the three
different techniques discussed in this paper, applied to the AR 8210
vector magnetogram sequence.  The background image in all three cases is
$\partial B_z / \partial t$.  The arrows in the upper left panel show
values of $\vecE_h^I$ from the PTD algorithm with $- \grad \psi$
set to zero.  
The top middle panel shows
$\vecE_h = \vecE_h^I - \grad_h \psi$, with $\psi$ computed using the
iteration technique.  
The top right panel shows $\vecE_h$ computed with the
variational technique (minimizing $E^2$).  
The bottom three panels show $E_z^I$ (left), 
$E_z^I - \partial \psi / \partial z$
computed with the iteration technique (middle),  and 
$E_z$ from
the variational method (right). 
Note that $E_z$ is poorly behaved
in the latter two cases along magnetic neutral lines.
All of the electric field vectors are drawn at the same scale, and the
$E_z$ images are displayed with the same linear grayscale colormap.
}
\label{figure:comparison}
\end{figure}

We have also tested the iteration technique and the variational technique
on the AR 8210 vector magnetogram data.  Figure \ref{figure:comparison} shows
a comparison of $\vecE$ computed using the three different techniques.  The
two left panels show $\vecE_h$ and $E_z$ from the PTD solution, with the
boundary conditions as described above.  The two middle panels show the same
quantities from the iterative technique, and the two right panels show the
solution from the variational method.  

There are common patterns in $\vecE_h$
seen in all three solutions, namely that $\vecE_h$ swirls around a region of
decreasing positive 
$B_z$ on the right hand side of the magnetogram about 40\% of
the way up from the bottom, and all three
show a common pattern near the large sunspot.
But the PTD solution also shows clear evidence of
artifacts as well, such as a strong
horizontal electric field normal to the magnetogram boundary near the
left edge that is non-existent in the variational solution, and much less
pronounced in the iteration solution.  
The iterative technique shows that
some of the artifacts in the PTD solutions are reduced, but also shows
large amplitude signals in some of the weak field regions.  
The variational solution shows small
electric field vectors in regions of small magnetic field strength, which
seems physical, at least superficially.  
There is little resemblance between the $E_z$ solution found with PTD, and the
$E_z$ maps of the iterative and variational methods.  Most likely, this is
because the latter two methods enforce $\vecE \cdot \vecB = 0$, meaning that
$E_z$ must be adjusted during solution algorithm
such that this condition is satisfied.
The iterative and variational solutions for $E_z$ are poorly constrained 
near polarity inversion lines, judging by the large fluctuations near them,
probably also a consequence of forcing
$\vecE \cdot \vecB=0$.  This might indicate that evolution is not consistent
with ideal MHD along polarity inversion lines.

\begin{figure}
\includegraphics[width=6.5in]{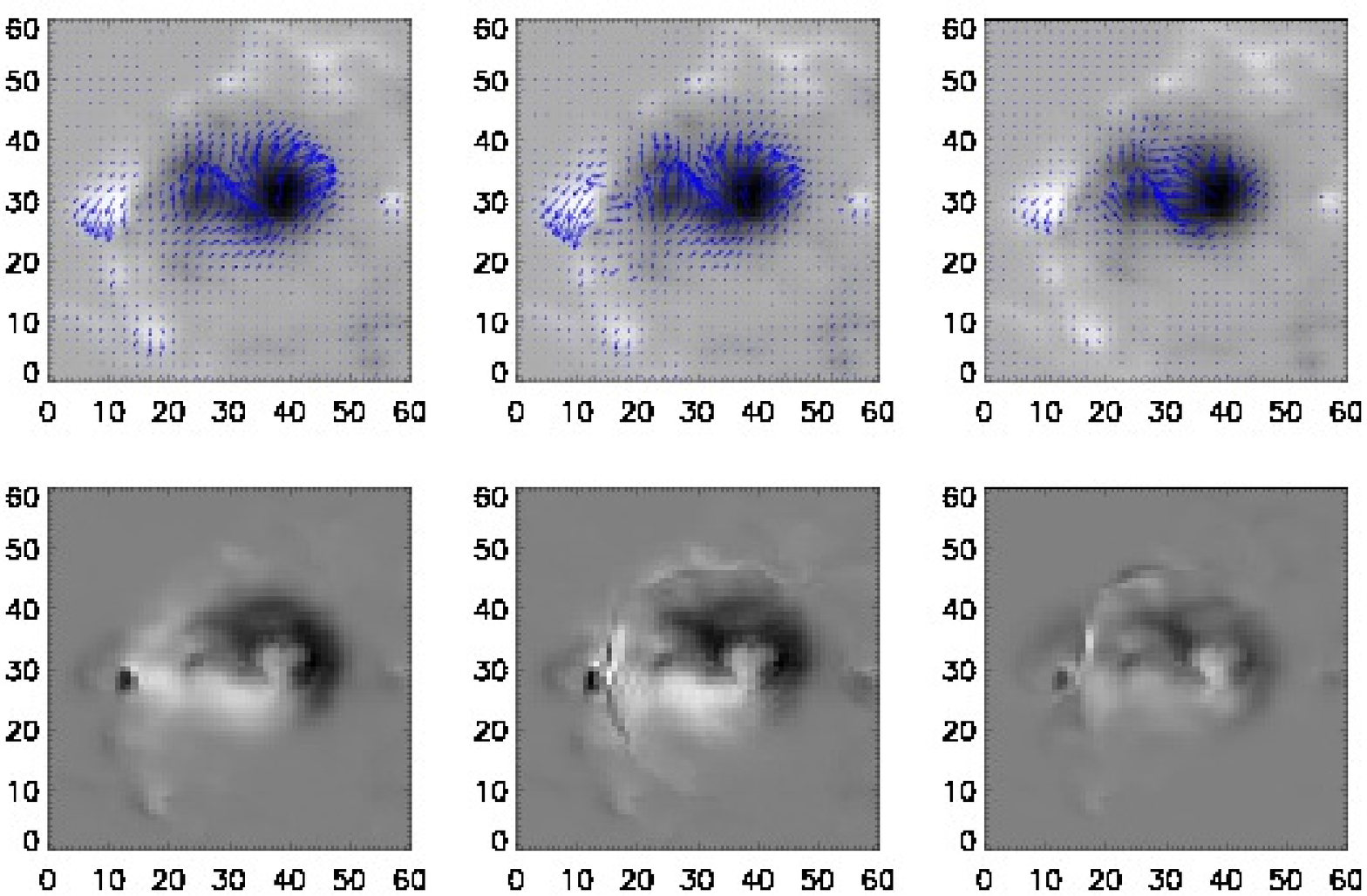}
\caption{
The top three panels show the derived horizontal Poynting flux using the
different techniques discussed in this paper, applied to the AR 8210
vector magnetogram sequence.  The background image in all three cases is
$B_z$.  The arrows in the upper left panel show
values of $\vecS_h^I$ from the PTD algorithm with $- \grad \psi$ contribution
set to zero.  
The top middle panel shows
$\vecS_h$ computed from $\vecE^I_h - \grad \psi$
using the iteration technique.  
The top right panel shows $\vecS_h$ computed with the
variational technique (minimizing $E^2$).  
The bottom three panels show $S_z^I$ (left), 
$S_z$
computed with the iteration technique (middle),  
and 
$S_z$ from
the variational method (right).  The vectors in the top three panels, and the
grayscale images in the bottom three 
panels are all plotted with the same scale. 
For a better view of the Poynting flux structure, only a 60x60 pixel region 
of the vector magnetogram field of view is shown.
}
\label{figure:poynting_8210}
\end{figure}

Finally, we show the Poynting fluxes found from all three electric field 
techniques in Figure \ref{figure:poynting_8210}.  
In contrast to the electric fields
themselves, the Poynting flux distributions are all fairly similar, showing
a northward (positive in the $\vecyhat$ direction) flux of magnetic energy
out of the negative polarity sunspot, and positive vertical Poynting flux
in the region south of the negative sunspot.  


\section{Summary and Discussion}
\label{section:conclusions}
In this paper, we introduce and demonstrate three new techniques for
determining electric field distributions given a sequence of vector
magnetogram observations.  The following is a summary of the most 
important points.

The first technique, based on a poloidal-toroidal decomposition (PTD) of the
time derivative of $\vecB$ and hence $c \grad \times \vecE$, 
results in a solution 
for $\vecE$ that satisfies all three components of Faraday's law, but does
not constrain any components of $\vecE$ derived from the gradient of
a potential $\psi$.
We showed in \S \ref{subsection:nopsi} and in \S \ref{section:8210} that
the unmodified PTD solutions (assuming $\grad \psi = 0$)
result in several artifacts and non-physical effects,
and argued that more physically realistic solutions for $\vecE$ must include
contributions to the electric field from a potential function, $-\grad \psi$.
One concern in particular is that the PTD solution has significant components
of $\vecE$ parallel to $\vecB$, contradicting the ideal MHD assumption that
many regard as most likely to represent the solar photosphere, and certainly
contradicting the ANMHD simulation test case discussed in detail throughout
\S \ref{section:psi}.  In spite of these defects, the electric fields found
from the PTD solutions were more accurate in the strong magnetic field
regions of the ANMHD simulation than those determined from all but one of the 
velocity inversion techniques tested in \cite{Welsch2007}, and were
as accurate as those of the MEF method.
The PTD solutions for $\vecE^I$ also
enable easy-to-compute estimates of Poynting and helicity fluxes.

We then explored two techniques for computing a potential function $\psi$.  
First, in \S \ref{subsection:iterative}, we outlined an iterative
procedure for computing a solution $-\grad \psi$ that acts to cancel the
components of $\vecE$ parallel to $\vecB$.  This technique looks promising
when applied to the ANMHD test case, significantly reducing scatter
of the inverted electric field values with those from the ANMHD simulations.
When compared only in the strong field regions analyzed by \cite{Welsch2007},
the iteration method performed at a similar level to the PTD solutions.  The
improvement to the solutions seen in the scatter plots evidently occurs
mainly in the weaker magnetic field regions.
The solutions derived from this method are not unique, however,
and when
applied to the NOAA AR 8210 example (\S \ref{section:8210}),
resulted in large electric field values
in regions of small magnetic field, which would imply unphysically large
velocities if an ideal MHD interpretation is assumed.

Second, in \S \ref{subsection:variational} we explored a variational approach to
computing $\psi$, by demanding that the total electric field minimize a
positive definite integral over the magnetogram field of view, and that it
obey the constraint $\vecE \cdot \vecB=0$ (though the formalism allows for
relaxation of this constraint).  This is essentially the same concept as
Longcope's Minimum Energy Fit (MEF) 
technique \citep{Longcope2004}.  However,
we extend his technique in two important ways:  first, the minimization
integral is generalized beyond the kinetic energy case considered by 
Longcope; second, we discovered that the variational solution for $\psi$ can be
combined with $\vecE^I$, resulting in a single equation
for a new scalar function $\chi$ whose gradient is proportional
to the horizontal Poynting flux.  The total electric field, including both
the inductive and potential contributions, is then computed {\it post facto}
from $\grad \chi$.  One can find $\grad \psi$, if desired, by
subtracting the PTD solutions $\vecE^I$
from the full electric field solutions derived from $\grad \chi$.

We applied the variational
technique to the ANMHD simulation results, and to the AR8210
vector magnetogram pair.  The variational method does a poor job of 
reproducing the electric fields in the ANMHD simulation, at least
in the one case we tried of minimizing $E^2$ ($i.e.$, setting $W^2=1$).  
On the other hand, the
solutions do correctly solve the induction equation, and they do so with
a much smaller electric field than the PTD or iteration method solutions,
or even the actual electric fields from the ANMHD simulation.
The variational technique applied to AR8210 shows large electric fields
mainly in regions where the magnetic field is changing rapidly, with the
electric fields elsewhere being very small.  
Both the PTD and iterative solutions
show large electric field vectors in regions of insignificant magnetic field.
In this case, the variational solution 
seems more physical, though this is a subjective evaluation.

We conclude that
with only the magnetic field measurements, it simply is not possible to
recover the true electric field in the solar photosphere -- 
too much information is missing, since
the true electric field depends not only on the induction equation, but also
on solutions of the momentum and energy
equations, about which vector magnetograms provide no direct information.
But it {\it is} possible to find an electric field solution that is 
both physically reasonable, and consistent 
with the observed evolution of $\vecB$.
Our results thus far
indicate that the variational technique does minimize electric field artifacts
in regions where we don't expect significant electric fields.

The PTD formalism for $\vecB$ leads to a useful and easy decomposition of
the observed magnetic field into potential and non-potential contributions
to the field.
One appealing application of this result
is that it suggests a recipe for evolving magnetograms as bottom
boundaries of MHD simulations, from potential
field distributions toward the actual magnetic field distribution.  By
starting from an initial potential field model in which $\scrJ=0$ (see
Appendix \ref{app:a}), and
then constructing a synthetic time evolution $\dot \scrJ$ that evolves toward
the observationally determined distribution of $\scrJ$ from the static
PTD solutions, one could impose $c E_z = - \dot \scrJ$
(along with $\vecE_h=0$) and allow an MHD simulation to evolve a solar
atmospheric model toward the observed magnetogram state.  It is important
to emphasize, however, this is generally not consistent with an ideal MHD
model for the photospheric 
electric field -- one would need to have an MHD code with
the flexibility to accommodate a user-specified electric field 
at the simulation boundary.
  
The iteration and variational solutions, since they explicitly enforce
$\vecE \cdot \vecB=0$, could be used both to define velocity fields at the
photospheric boundary, and for 
assimilating a time series of vector magnetograms
directly into MHD models
\citep{Abbett2010}.  In the
future, these solutions will be more thoroughly 
compared and contrasted with the more
conventional inductive and tracking-based solutions for the velocity
field from vector magnetogram data.

%
%
%
%
%
%
%
%
\acknowledgments We gratefully acknowledge funding from the
Heliophysics Theory Program, under NASA grant NNX08AI56G-04/11,
as well as support from the NASA LWS TR\&T award NNX08AQ30G.  
This work was also supported by NSF SHINE award ATM-0752597, NSF
SHINE award ATM-0551084, and NSF support to our group at SSL for the 
University of Michigan's CCHM project, ATM-0641303.
We thank Dana Longcope
for his discussions about the MEF technique, and for verifying our result
that MEF implies a curl-free condition on 
$ 1 / ( B^2 B_z )\ \vecE \times \vecB$.
We thank Peter MacNeice for pointing out that the degeneracy of the
equations for $\partial \scrB / \partial z$ and $\scrJ$ at the boundaries
were solutions of the homogeneous Cauchy-Riemann equations.  
One of the
authors is grateful to Bombay Sapphire Gin for providing the insight
that a PTD solution for the electric field from vector magnetogram data
was possible.

\newcounter{appendix}
\begin{appendix}
\setcounter{appendix}{0}
\section{Potential Magnetic Fields Described with the PTD Formalism}
\label{app:a}

The PTD formalism allows for an alternative approach for describing potential
magnetic fields near the magnetogram layer.  
Normally, potential magnetic fields derived from 
magnetograms use the normal component of the field on the photospheric
boundary, plus boundary conditions for the magnetic field at the side
and upper boundaries (sometimes taken to be at infinity)
to derive a solution within a specified volume.
The horizontal fields of the potential solution
on the bottom boundary are then determined by taking the horizontal gradient
of the scalar potential that describes the potential field.  The horizontal
components of the potential field at the photosphere thus depend indirectly
on the assumed behavior of the field at a significant distance from the
photosphere.  In this Appendix, we show how the horizontal
field components in a vector magnetogram can be used to define a 
potential field solution using the PTD formalism, as an alternative to
using an assumed behavior of the field at distant boundaries.

Using equation (\ref{equation:ptdb}) and setting all three components of
the electric current to zero, one can show that
$\nabla_h^2 \scrJ = 0$ and $\nabla_h^2\ (\nabla^2 \scrB )=0$. 
This condition can be met with $\scrJ$ and $\nabla^2 \scrB$ being functions
of $z$ only.  We will assume henceforth that both functions of $z$
are equal to the special case of zero:
\be
\nabla^2 \scrB = \scrJ = 0
\label{equation:specialcasezero}
\ee
Since a potential magnetic field can be specified with 
a single potential function, we assume we
that we can find a single $\scrB$ function that can represent an
arbitrary potential field, and ignore $\scrJ$. 

What is the relationship between the usual scalar potential function $\phi$
($\vecB = - \grad \phi$) and the corresponding $\scrB$ function for the same
potential field?  To distinguish between the general case and the 
potential-field case, we denote the potential field case of $\scrB$ 
as $\scrB_P$.  Further, here we consider $\scrB_P$ to be an explicit
function of 
three-dimensional space, in contrast to the convention in the rest of the 
paper that the scalar potentials are assumed to
depend only on the two-dimensional domain 
of the vector magnetogram.  From equation
(\ref{equation:ptdb}) without the contribution from $\scrJ$,
\be
\vecB = \grad \times \grad \times \scrB_P \veczhat\ .
\label{equation:ptdpot}
\ee
If the field is potential and thus current-free, then 
from equation (\ref{equation:specialcasezero})
$\scrB_P$ satisfies
the three dimensional Laplace equation
\be
\nabla^2 \scrB_P = 0\ .
\label{equation:laplacepot}
\ee
In the PTD formalism, the horizontal field on the plane of the magnetogram
is given by
\be
\vecB_h = \grad_h {\partial \scrB_P \over \partial z}
\label{equation:bhpot}
\ee
while the vertical field is given by
\be
\nabla_h^2 \scrB_P\ = - B_z\ .
\label{equation:bzpot}
\ee
However, since $\scrB_P$ obeys the three-dimensional 
Laplace equation, the left hand side of
equation (\ref{equation:bzpot}) is also equal to 
$- {\partial^2 \scrB_P} / {\partial z^2}$, and therefore, the magnetic field
can be written as
\be
\vecB = - \grad \left ( {- \partial \scrB_P \over \partial z} \right )\ ,
\label{equation:potboth}
\ee
meaning that one can then make the identification
\be
\phi = - {\partial \scrB_P \over \partial z}\ .
\label{equation:phiscrb}
\ee

The PTD formalism allows one to use two different
parts of the observed data in computing properties of the potential field.  
As noted above,
the conventional approach uses the observed normal component ($B_z$)
of the field at the magnetogram as the only photospheric boundary condition.  
The PTD formalism allows one to include the divergence of the horizontal
field in the solution, as well as the normal component of $\vecB$.
Equation (\ref{equation:poissondivbh}), applied to $\scrB_P$, shows that
\be
\nabla_h^2 \left ( {\partial \scrB_P \over \partial z} \right ) 
= \grad_h \cdot \vecB_h
= - {\partial B_z \over \partial z}
\label{equation:bpotgrad}
\ee
This uses the horizontal magnetic field data to specify the 
rate at which $B_z$ 
decreases immediately above the surface of the magnetogram in the potential
field solution.  To make the solution of the two-dimensional Poisson equation 
(\ref{equation:bpotgrad}) well-posed,
one can apply the Neumann boundary condition at the edge of the magnetogram,
\be
{\partial \over \partial n} 
\left ( {\partial \scrB_P \over \partial z} \right ) = B_n\ , 
\label{equation:bdrybpot}
\ee
where as in \S \ref{subsection:bdecomp}, $B_n$ is the observed
component of $\vecB_h$ normal to the magnetogram boundary.  

From the
solution to the horizontal Poisson equation 
(\ref{equation:bpotgrad}) for
$\partial \scrB_P / \partial z$,
one can then use equation (\ref{equation:bhpot}) to 
reconstruct the contribution 
to $\vecB_h$ that can be 
ascribed solely to a potential magnetic field, without having to assume
any behavior at distant boundaries.  On the other hand, using the potential
field derived in this way to compute the behavior far from the magnetogram
is probably dangerous, as any errors in the field measurement are likely to
be greatly magnified when extrapolated to large distances.

Finally, it is frequently useful to be able 
to express the potential field in terms
of a vector potential $\vecA_P$, rather than a scalar potential.  
Estimates of
the magnetic helicity flux through the photosphere typically involve
$\vecA_P$.
Using the PTD formalism, the vector potential
$\vecA_P$ for the potential field is found
simply from equation (\ref{equation:vecp}):
\be
\vecA_P = \grad_h \times \scrB_P \veczhat\ ,
\label{equation:vecpa}
\ee
evaluated in the plane of the magnetogram, and where $\scrB_P$ 
is found from a solution
to the two-dimensional Poisson equation (\ref{equation:bzpot}).  
Note that $\vecA_P$ has no vertical ($z$) components.  


\refstepcounter{appendix}
\section{Fourier Transform Solutions to the PTD equations}
\label{app:b}

If the boundary conditions for the vector magnetogram are assumed to be
periodic, and if the magnetic field's time derivative does not have
a uniform (zero-wavenumber) component,
Fast Fourier Transform (FFT) techniques greatly simplify the solutions for the 
Poisson equations for $\dot \scrB$, $\dot \scrJ$, and 
$\partial \dot \scrB / \partial z$. 
If we denote the Fourier transforms of $\dot B_x$, $\dot B_y$, and $\dot B_z$ as
$\tilde{\dot B_x}$, $\tilde{\dot B_y}$, and $\tilde{\dot B_z}$, 
respectively, one can write the solutions to equations 
(\ref{equation:poissonbz}-\ref{equation:poissondivbh}) as
\be
\dot \scrB = \scrF^{-1}\left( \tilde{\dot B_z} \over k_x^2+k_y^2\right) ,
\label{equation:fftbdot}
\ee
\be
\dot \scrJ = \scrF^{-1}\left( i ( k_x \tilde{\dot B_y} - k_y \tilde{\dot B_x} ) 
\over k_x^2+k_y^2\right) ,
\label{equation:fftjdot}
\ee
and
\be
{\partial \dot \scrB \over \partial z} = \scrF^{-1}\left( 
{ - i ( k_x \tilde{\dot B_x} + k_y \tilde{\dot B_y} ) \over k_x^2+k_y^2 } ,
\right)
\label{equation:fftdbdzdot}
\ee
where $\scrF^{-1}$ denotes the inverse Fourier transform, and $k_x$ and $k_y$
are the horizontal wavenumbers in the $x-$ and $y-$ 
directions within the magnetogram,
respectively.  The quantities $\dot \scrB$ and 
$\dot \scrJ$ can then be used to derive
the inductive electric field $\vecE^I$ via equation (\ref{equation:eptd}).

The same approach 
can be used to find the functions $\scrB$, $\scrJ$, and 
$\partial \scrB / \partial z$ from Fourier transforms of the magnetic field
$\tilde B_x$, $\tilde B_y$, and $\tilde B_z$ by using the same equations
(\ref{equation:fftbdot}-\ref{equation:fftdbdzdot}) 
except with the Fourier transforms 
of the time derivatives of the magnetic
field components replaced with the Fourier transforms of the 
magnetic field components themselves.  

However, if the magnetic field, or its
time derivative, has a non-zero average value
in any of the component directions, the FFT solutions cannot account for this,
and one must use the techniques described in Appendix \ref{app:c} to correct
the FFT-derived solution.
Equation (\ref{equation:vecp}) can 
then be used to to
find the vector potential (ignoring the gauge contribution).  

\refstepcounter{appendix}
\section{Accounting For Average Values of the Magnetic Field 
and its Time Derivative}
\label{app:c}

If the magnetic field time derivative contains a spatially 
uniform (zero-wavenumber) component $\dot \vecB_0$, 
equations (\ref{equation:fftbdot}-\ref{equation:fftjdot}) 
will not recover this component, because the assumption
of periodic boundary conditions for $\vecE$ cannot produce a uniform vector
$\grad \times \vecE$ and hence $\dot \vecB$.  An additional electric field
component must be explicitly added to account for the uniform time
derivative of $\vecB$.  Similarly, if Neumann boundary conditions are
used to solve equation (\ref{equation:poissonbz}), the solution will
force an assumption that the spatial average of $\dot B_z$ is zero, and
an additional term must be added to the derived solution for $\vecE$.

A term of the form
\be
c \vecE_0 = -  {1 \over 2}  \dot \vecB_0 \times ( \vecr -\vecr_0 ) ,
\label{equation:uniformbdot}
\ee
will fully reproduce the observed time derivative of the magnetic field
when added to the electric field derived by using equations 
(\ref{equation:fftbdot}-\ref{equation:fftjdot}) in equation
(\ref{equation:eptd}).  Here $\vecr$ is the 
position vector, and $\vecr_0$ can be any constant vector offset.

To correct equation (\ref{equation:curleptdexpand}) for $c \grad \times \vecE$, 
it is sufficient to just
add the term $- \dot \vecB_0$ to equation (\ref{equation:curleptdexpand}) 
if using the FFT-derived solutions
for $\dot \scrJ$ and $\dot \partial \scrB / \partial z$ given in Appendix
\ref{app:b}. 

A similar problem arises if the magnetic field itself has a non-zero
uniform component $\vecB_0$: 
the Fourier transform-derived solutions for $\vecA$ 
will not be able to recover $\vecB_0$.  
Instead, a term of the form
\be
\vecA_0 = {1 \over 2}\ \vecB_0 \times \left( \vecr - \vecr_0 \right)
\label{equation:uniformb}
\ee
when added to equation (\ref{equation:vecp}) will reproduce the full
magnetic field observation as $\grad \times \vecA$.  Note that
$\vecE_0 = {-1}/{c}\  {\partial \vecA_0}/ {\partial t}$.

The solutions for $c \vecE_0$ depend on an unspecified position vector offset,
$\vecr_0$.  We now present an argument for determining $\vecr_0$,
based on the concept of Galilean invariance.

If the electric field contribution from equation (\ref{equation:uniformbdot}) 
originates from an ideal electric field due 
to a velocity field $\vecv_0$ in the presence of the uniform magnetic field
component $\vecB_0$, then we can equate the
two expressions for the electric field:
\be
-{1 \over 2c} \dot \vecB_0 \times ( \vecr - \vecr_0 ) = 
- { \vecv_0 \over c} \times \vecB_0 .
\label{equation:Galilean_start}
\ee
Taking the cross-product of both sides of equation 
(\ref{equation:Galilean_start}) with the vector ${\dot \vecB}_0$, 
one finds after some manipulation
\be
{1 \over 2} \dot B_0^2 { (\vecr - \vecr_0 ) }_{\perp} = 
(\vecv_0 \times \vecB_0 ) \times 
\dot \vecB_0 ,
\label{equation:Galilean_second}
\ee
where $\dot B_0$ is the amplitude of $\dot \vecB_0$, and subscript $\perp$
denotes the directions perpendicular to $\dot \vecB_0$.  Taking a spatial
average of equation (\ref{equation:Galilean_second}) then results in
\be
{1 \over 2} \dot B_0^2 { (\bar \vecr - \vecr_0 ) }_{\perp} = 
(\bar \vecv_0 \times \vecB_0 ) \times \dot \vecB_0 .
\label{equation:Galilean_third}
\ee
Now we can identify the spatial average $\bar \vecv_0$ 
with $\vecv_{ref}$, the velocity of a Galilean reference frame.  
In other words, if two magnetograms in a sequence 
have an overall 
non-zero net shift, due to $e.g.$ an inaccurate (or non-existent) 
correction for solar rotation,
we then identify the overall reference frame velocity responsible for
this shift as $\vecv_{ref}$.  If the two magnetograms have been co-registered
such that the net overall shift is zero, then we assume that $\vecv_{ref} = 0$.
In that case, the right hand side of equation (\ref{equation:Galilean_third})
must be zero, and one then finds this constraint for $\vecr_0$:
\be
{\vecr_0}_{\perp} = {\bar  \vecr  }_{\perp} .
\label{equation:Galilean_last}
\ee
This condition is always 
satisfied when the vector offset $\vecr_0$, which can also be regarded as the
origin of the vector magnetogram
coordinate system, is chosen to coincide with the geometric center 
of the vector magnetogram $\bar \vecr $.  By determining $\vecr_0$, 
this allows us 
to determine an electric field and vector potential solution unambiguously.
If one wants to perform the calculation in a moving reference frame with
a non-zero $\vecv_{ref}$, the value of $\vecr_0$ will then be determined from
equation (\ref{equation:Galilean_third}) instead 
of equation (\ref{equation:Galilean_last}).

It is useful to evaluate the result after applying this
substitution into equations (\ref{equation:uniformbdot}) 
and (\ref{equation:uniformb})
in component
form for a vector magnetogram lying on the surface $z=0$.  In that case,
$\bar x$ and $\bar y$ represent the average values of $x$ and $y$ at the center
of the magnetogram, and $\bar z = 0$.  We find
\bea
c \vecE_{0} = {1 \over 2} [ (y - \bar y) \dot B_{0,z} 
- (z - \bar z) \dot B_{0,y} ] \vecxhat 
\nn
\\
+{1 \over 2} 
[- (x - \bar x ) \dot B_{0,z} + (z - \bar z) \dot B_{0,x} ] \vecyhat 
\nn
\\
+ {1 \over 2} [ (x - \bar x) \dot B_{y,0} 
- (y - \bar y) \dot B_{x,0} ] \veczhat \ ,
\label{equation:Galilean_E}
\eea
and
\bea
\vecA_{0} = {1 \over 2}  [ - (y - \bar y) B_{0,z} 
+ (z - \bar z) B_{0,y}] \vecxhat 
\nn
\\
+ {1 \over 2}[ (x - \bar x ) B_{0,z} - (z - \bar z) B_{0,x} ] \vecyhat 
\nn
\\
+ {1 \over 2} [ - ( x - \bar x) B_{y,0} + (y - \bar y) B_{x,0} ] \veczhat \ ,
\label{equation:GalileanA}
\eea
where $\dot B_{0,x}$, $\dot B_{0,y}$, and $\dot B_{0,z}$ 
are the components of $\dot \vecB_0$, and $B_{0,x}$, $B_{0,y}$, and $B_{0,z}$
are the components of $\vecB_0$.  
While the terms proportional to $(z - \bar z)$ will vanish at the
photosphere in $\vecA_0$ and $c \vecE_0$, they will still contribute to
vertical derivatives of $\vecA_0$ and $c \vecE_0$ evaluated at the
photosphere.


%
%
%

\refstepcounter{appendix}
\section{Deriving the Variational Equation for $\chi$}
\label{app:cd}
Equation (\ref{equation:chidef}) can be re-written as
\be
( W^2 / B_z ) ( c \vecE_h \times B_z \veczhat + c E_z \veczhat \times \vecB_h )
= - \grad_h \chi ,
\label{equation:chidefbreakout}
\ee
or
\be
c \vecE_h \times \veczhat + ( c E_z / B_z ) \veczhat \times \vecB_h = - 
{ \grad_h \chi \over W^2} .
\label{equation:chidefoverw2}
\ee
Taking the dot product of equation (\ref{equation:chidefoverw2}) with
$\veczhat \times \vecB_h$, and making use of equation (\ref{equation:jgone})
namely
\hbox{$E_z B_z = ( \vecR \cdot \vecB -\vecE_h \cdot \vecB_h )$}, we find
\be
{ c E_z } =  B_z {c \vecR \cdot \vecB \over B^2} - 
B_z {\grad_h \chi \over W^2 B^2} \cdot {\veczhat \times \vecB_h} .
\label{equation:ezoverbz}
\ee
Substituting equation (\ref{equation:ezoverbz}) into equation 
(\ref{equation:chidefoverw2}) then results, after some manipulation, in
\be
c \vecE_h \times \veczhat + {c \vecR \cdot \vecB \over B^2}\ \veczhat 
\times \vecB_h + {1 \over W^2 B^2}  ( \vecB_h \cdot \grad_h \chi ) \vecB_h =
-{B_z^2 \over W^2 B^2 } \grad_h \chi
\label{equation:chivec}
\ee
At this point, the only quantity not involving $\chi$ that is unknown in
equation (\ref{equation:chivec}) is $\vecE_h$ (recall that $\vecR$ is 
assumed to be either zero or known {\it a priori}).  If we take the
horizontal divergence of equation (\ref{equation:chivec}), 
however, we can eliminate $c \vecE_h$ through the use of 
the magnetic induction equation
$\grad_h \cdot ( c \vecE_h \times \veczhat ) = 
c \veczhat \cdot \grad_h \times \vecE_h = - \partial B_z / \partial t $
resulting in
\be
{- \partial B_z \over \partial t} + \grad_h \cdot 
\left( {c \vecR \cdot \vecB \over B^2}\ \veczhat
\times \vecB_h \right) + \grad_h \cdot \left( {1 \over W^2 B^2}  
( \vecB_h \cdot \grad_h \chi ) \vecB_h \right) = - \grad_h \cdot 
\left( {B_z^2 \over W^2 B^2 } \grad_h \chi \right)  . 
\label{equation:chidiv}
\ee

\end{appendix}


\end{document}